\documentclass[sigconf]{acmart}
\AtBeginDocument{%
  }

\copyrightyear{2026}
\acmYear{2026}
\setcopyright{cc}
\setcctype{by}
\acmConference[CHI '26]{Proceedings of the 2026 CHI Conference on Human Factors in Computing Systems}{April 13--17, 2026}{Barcelona, Spain}
\acmBooktitle{Proceedings of the 2026 CHI Conference on Human Factors in Computing Systems (CHI '26), April 13--17, 2026, Barcelona, Spain}
\acmDOI{10.1145/3772318.3790999}
\acmISBN{979-8-4007-2278-3/2026/04}

\usepackage{enumitem}
\usepackage{graphicx}
\usepackage{caption}
\usepackage{subcaption}
\newcommand{\sysName}{AnkleType}

\begin{document}

%%
%% The "title" command has an optional parameter,
%% allowing the author to define a "short title" to be used in page headers.
\title{\sysName: A Hands- and Eyes-free Foot-based Text Entry Technique in Virtual Reality}

\author{Xiyun Luo}
\orcid{0009-0005-8741-5755}
\affiliation{%
  \institution{Department of Computer Science\\ Shantou University}
  \city{Shantou}
  \country{China}}
\email{24xbluo@stu.edu.cn}

\author{Weirong Luo}
\orcid{0009-0002-4587-3617}
\affiliation{%
  \institution{Department of Computer Science\\ Shantou University}
  \city{Shantou}
  \country{China}}
\email{24wrluo@stu.edu.cn}

\author{Kening Zhu}
\orcid{0000-0001-6740-4921}
\authornote{He is also with City University of Hong Kong Shenzhen Research Institute, Shenzhen, China.}
\affiliation{%
  \institution{School of Creative Media\\ City University of Hong Kong}
  \city{Hong Kong}
%   \streetaddress{Run Run Shaw Creative Media Centre, Level 7, 18 Tat Hong Avenue, Kowloon Tong}
  \country{China}}
\email{keninzhu@cityu.edu.hk}

\author{Taizhou Chen}
\orcid{0000-0002-7005-4560}
\authornote{Corresponding author.}
\affiliation{%
  \institution{Department of Computer Science\\ Shantou University}
  \city{Shantou}
  \country{China}}
\email{taizhou.chen@my.cityu.edu.hk}

%%
%% By default, the full list of authors will be used in the page
%% headers. Often, this list is too long, and will overlap
%% other information printed in the page headers. This command allows
%% the author to define a more concise list
%% of authors' names for this purpose.

\newcommand{\revise}[1]{{\color{black}#1}}
\newcommand{\newrevise}[1]{{\color{black}#1}}

\renewcommand{\shortauthors}{Luo et al.}

%%
%% The abstract is a short summary of the work to be presented in the
%% article.
\begin{abstract}
  Virtual Reality (VR) emphasizes immersive experiences, while text entry often requires hands or visual attention, which may disrupt the interaction flows in VR. We present \sysName, a hand- and eye-free text-entry technique that leverages ankle-based gestures for both standing and sitting situations. We began with two preliminary studies: one investigated the movement range of users' ankles, and the other elicited user-preferred ankle gestures for text-entry-related operations. The findings of these two studies guided our design of \sysName. %These findings guide our exploration of ankle-based typing. 
  To optimize \sysName's~ keyboard layout for eye-free input, we conducted a user study to capture the users’ natural ankle spatial awareness with a computer-simulated language test. %to reduce word ambiguity. 
  Through a pairwise comparison study, we designed a bipedal input strategy for sitting (\textbf{BPSit}) and a unipedal input strategy for standing (\textbf{UPStand}). Our first in-VR text-entry evaluation with 16 participants demonstrated that our methods could support the average typing speed 
  % up to 9.44 WPM 
  \revise{from 8.99 WPM (\textbf{BPSit}) to 9.13 WPM (\textbf{UPStand})}
  for our first-time users. We further evaluated our design with a 7-day longitudinal study with \revise{12} participants.
  \revise{Participants achieved an average typing speed of 15.05 WPM with \textbf{UPStand} and 16.70 WPM with \textbf{BPSit} in the visual condition, and 11.15 WPM and 12.87 WPM, respectively in the eyes-free condition.}
  % The results showed that the participants achieved eye-free input speeds of {\color{red}12.87} WPM with \textbf{BPSit} and {\color{red}11.15} WPM with \textbf{UPStand}, outperforming current hand-free typing techniques for VR.
\end{abstract}

%%
%% The code below is generated by the tool at http://dl.acm.org/ccs.cfm.
%% Please copy and paste the code instead of the example below.
%%

%%
%% Keywords. The author(s) should pick words that accurately describe
%% the work being presented. Separate the keywords with commas.
%% A "teaser" image appears between the author and affiliation
%% information and the body of the document, and typically spans the
%% page.
\begin{teaserfigure}
  \includegraphics[width=\textwidth]{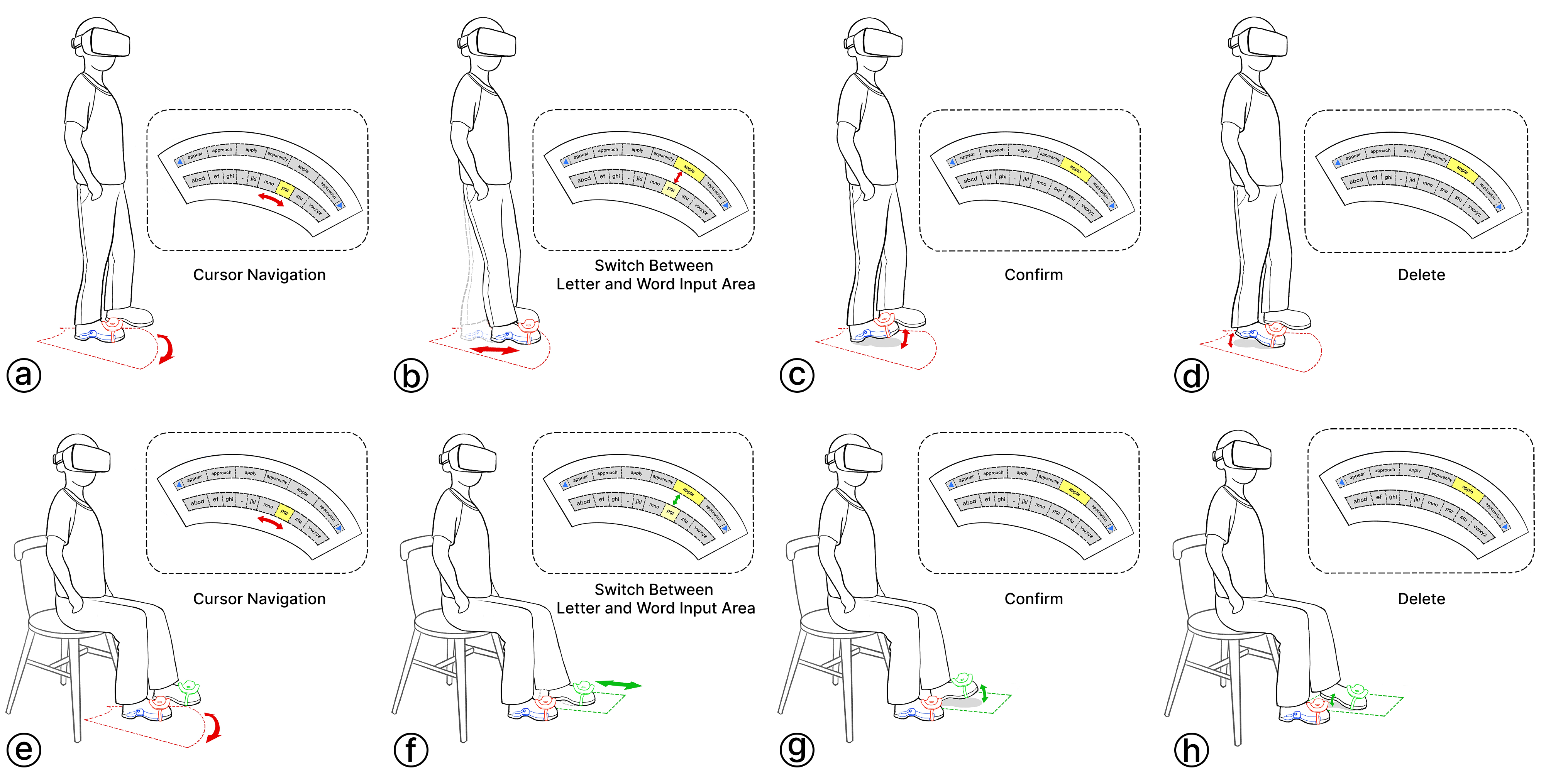}
  \Description{AnkleType enables eye- and hand-free foot-based typing for standing (UPStand a-b) and sitting posture (BPSit e-h). ae Rotate the ankle for cursor navigation, bf Foot flat forward and Foot flat backward gesture for switching between the letter input area and word input area, cg Forefoot single tap gesture to confirm the input, dh Rearfoot single tap to delete the inputted letter or word.}
  \caption{\sysName{} enable eye- and hand-free foot-based typing for standing (\textbf{UPStand} \textcircled{a} - \textcircled{d}) and sitting posture (\textbf{BPSit} \textcircled{e} - \textcircled{h}). \textcircled{a}\textcircled{e} Rotate the ankle for cursor navigation, \textcircled{b}\textcircled{f} \textit{Foot flat forward} and \textit{Foot flat backward} gesture for switching between the letter input area and word input area, \textcircled{c}\textcircled{g} \textit{Forefoot single tap} gesture to confirm the input, \textcircled{d}\textcircled{h} \textit{Rearfoot single tap} to delete the inputted letter or word. }
  \label{fig:teaser}
\end{teaserfigure}

% \received{20 February 2007}
% \received[revised]{12 March 2009}
% \received[accepted]{5 June 2009}

%%
%% This command processes the author and affiliation and title
%% information and builds the first part of the formatted document.
\maketitle

\section{Introduction}

% # background
% [VR is becoming popular]
Text entry serves as a fundamental activity in a computing system to support communication, inquiry, annotation, and documentation across a variety of tasks~\cite{card1983psychology}. As the rapid development of hardware capability and the increasing portability of head-mounted display (HMD) has moved Virtual Reality and Augmented Reality (VR/AR) toward everyday applications such as gaming~\cite{niu2025exploring,kim2023bubbleu,cmentowski2023never}, creativity~\cite{numan2025cocreatar}, remote collaboration~\cite{wei2025remotechess,gronbaek2024blended}, and social applications~\cite{hide2025closer,cummings2025self,lechappe2025understanding}, the ability to input text efficiently and comfortably in a VR/AR system becomes increasingly important. 
% [Text entry in VR is essential]
However, conventional methods such as physical keyboards suffer from \revise{the reduction of} typing performance when adapted to VR environments~\cite{grubert2018text}, highlighting the need to investigate more intuitive and expressive text entry techniques for the VR context. 

% # existing solution and problem
% [Research investigate various of VR text entry technique such as gaze, bard hand, controller]
To mitigate this gap, researchers and designers have explored a variety of VR text entry techniques, including the adaptation of handheld controllers~\cite{wu2024fanpad, leng2022efficient}, head gestures~\cite{yu2017tap, lu2020exploration}, hand gestures~\cite{fallah2023h4vr}, virtual keyboards~\cite{akhoroz2024poke, streli2024touchinsight}, and speech~\cite{adhikary2021text}. 
% [VR aims to provide immersive experience, but text entry may disrupt the experience flow, calling need of investigating natural and expressive text entry approaches to support hand- and eye-free text entry]
While these approaches have made significant progress to support text entry in VR, they face limitations when users' hands or visual attention are \revise{less available. For example, when a user is holding objects physically or virtually using a controller, their hands become less available for typing (Figure~\ref{fig:Scenarios}). }
% occupied with other essential tasks, often disrupting the flow of immersive experiences in VR environment. 
Head-based and eye-gaze interaction, in particular, are prone to induce motion sickness meanwhile causing eye fatigue, which may increase physical and cognitive burden during long-term usage \cite{shi2021virtual,wang2022real,merhi2007motion}. Beyond these approaches, recent research~\cite{wan2024exploration} has pioneeringly explored foot-based text entry techniques for VR, demonstrating the feasibility and potential of leveraging the feet as a text input channel in VR. While this method provides a hands-free typing experience with low visual burden, it still requires users' attention on staring at the keyboard layout during typing and requires a wide range of motion with the user's feet.
% , which may increase users' fatigue. \hl{[See if our results can support this statements]} 
Furthermore, their study primarily focused on the sitting condition, as they pointed out that users do not prefer typing for a long time while standing. However, for those applications that require a standing posture, such as gaming, it is inevitable to type while standing, highlighting the need to investigate an optimal typing interface for standing scenarios. 

In this paper, we propose \sysName{}, a novel foot-based text entry technique for hand- and eye-free text entry in VR. The system employs ankle rotations for navigation with forefoot and rearfoot taps to perform control actions such as confirm and cancel. \sysName{} is explicitly designed to support both standing and sitting postures, offering an efficient and natural typing experience with only minimal foot movement. To explore the design space of this technique, we focused on two key aspects: keyboard layout and interaction design, and investigated and evaluated them through a series of user studies. 

We first conducted a preliminary elicitation study to learn users’ preferences and performance for ankle-based foot interaction using unipedal and bipedal input in both standing and sitting postures. Based on the observation of differences in ankle rotation range between postures, we explored two separate keyboard layouts for standing and sitting postures, respectively. To this end, we run a user study with a series of computer-simulated typing tests to compare the performance of various design options, yielding one optimal layout for each posture that balances eye-free feet reachability and language disambiguation. Building on this, we conducted a second study to identify the most effective interaction mechanisms for each posture. Inspired by our preliminary elicitation study, we designed and compared four user-defined mechanisms. The results showed that \textbf{unipedal stand (UPStand)} and \textbf{bipedal sit (BPSit)} offered the best balance of typing efficiency, comfort, and usability, as the final designs for \sysName{}. The results also show that our methods could support the average typing speed 
% up to 9.44 WPM
\revise{from 7.80 WPM to 9.44 WPM}
for our first-time users. Finally, we run a 7-day longitudinal study
%with visual cues removed midway through the sessions
to test the system's learnability and the eye-free performance. \revise{For the visual condition, result shows that \sysName{} achieves an average typing speed of 15.05 WPM for \textbf{unipedal stand (UPStand)} and 16.70 WPM for \textbf{bipedal sit (BPSit)}, with a total error rate of 3.71\% and 2.48\%, respectively. For the blind condition, result shows that \sysName{} achieves an average typing speed of 11.15 WPM for \textbf{unipedal stand (UPStand)} and 12.87 WPM for \textbf{bipedal sit (BPSit)}, with a total error rate of 9.91\% and 8.80\%, respectively.} The results also show that \sysName{}  not only outperformed state-of-the-art hand-free VR text entry techniques, but also achieved competitive performance in the eye-free VR text entry task.

Our contributions are fourfold:
\begin{itemize}
\item We propose a novel ankle-based typing technique that enables hands- and eyes-free text entry in VR, supporting both standing and sitting postures.
\item We present an elicitation exploration that identifies users' preferences to perform ankle-based foot gestures in both standing and sitting conditions.
\item We perform a series of empirical explorations to optimize the keyboard layout and the interaction design of the proposed technique across postures.
\item We conduct a longitudinal user study to demonstrate the learnability and the effectiveness of the proposed technique.
\end{itemize}

\section{Related Works}

\sysName{} is largely inspired by existing works on immersive text entry techniques and foot-based interfaces. In addition, we also refer to existing eye-free text entry techniques. 

\subsection{Text Entry for Immersive Environment}

Text entry in immersive virtual environments has been explored for decades \cite{bowman2002text}. 
Previous research reveals that text entry in immersive head-mount display using a conventional keyboard \cite{grubert2018text} suffers from the low entry speed and high error rate problem. To mitigate this gap, researchers have explored many novel text entry techniques, including handheld controller \cite{wu2024fanpad, leng2022efficient}, head gesture \cite{yu2017tap, lu2020exploration}, hand gesture \cite{fallah2023h4vr}, and virtual keyboard \cite{akhoroz2024poke, streli2024touchinsight}. While the solutions are diverse, we summarized and discussed them from two aspects: controller-based and non-controller-based. 

\subsubsection{Controller-Based Techniques}

Type with VR hand-held controller is one of the straightforward and popular input strategies, attracting various previous explorations.
% device due to their stability and integration with head-mounted displays. 
Among them, typing through controller ray casting is the most common input method, but it is less efficient and prone to increasing physical tedium during long-term usage. 
% To address this problem, Wan et al.\cite{wan2024design} proposed to improve security by randomizing layout and spacing. 
Therefore, prior research has investigated supporting more expressive and efficient text entry methods using controller motion, such as pointing or tapping. Speicher et al. \cite{speicher2018selection} explored a variety of VR text entry techniques, including a method by reversing and holding the HTC Vive controller for tapping, achieving an input speed of 12.96 WPM.
% Speicher et al.\cite{speicher2018selection} achieved a controller tapping input speed of 12.69 wpm. 
Leng et al. \cite{leng2022efficient} redesigned the keyboard layout for controller typing by arranging letters in a flower-shaped layout to reduce the controller pointing movement. Their method achieved an impressive typing performance of 22.97 WPM after a 6-day practice. 
% mapped QWERTY translation to a flower layout and achieved 22.97 wpm after practice. Controller solutions are diverse and efficient, but they occupy the user's hands, which becomes an obstacle in many scenarios that require hand interaction.

Beyond using controller motions, prior research has leveraged joysticks and touchpads on a conventional VR hand-held controller to support a variety of text entry techniques. PizzaText~\cite{yu2018pizzatext} uses dual joysticks to input on a "pizza" layout, with a speed of 15.85 WPM. HiPad~\cite{jiang2020hipad} designs a six-key radial layout on the touchpad, achieving an input speed of 13.57 WPM with single-handed input. FanPad~\cite{wu2024fanpad} maps QWERTY to dual touchpads to support rapid text entry with subtle motions. By optimizing the thumb trajectory, their method achieves a text entry speed of 19.73 WPM. 

% In addition, spatial input and tapping input also show more possibilities. Speicher et al.\cite{speicher2018selection} achieved a controller tapping input speed of 12.69 wpm. Flower Text Entry\cite{leng2022efficient} mapped QWERTY translation to a flower layout and achieved 22.97 wpm after practice. Controller solutions are diverse and efficient, but they occupy the user's hands, which becomes an obstacle in many scenarios that require hand interaction.

\subsubsection{Non-Controller-Based Techniques}

Numerous researchers have dedicated themselves to exploring non-controller-based text entry techniques, such as using hand or head gestures. 
% Non-controller input technology aims to get rid of dependence on controllers and pursue a more natural interactive experience. 
Mid-air typing, as one of the intuitive text entry strategies in immersive environments that adapt to users' traditional behaviors, has been widely explored by the research community. However, although natural, air typing is suffering from a problem of lacking support and tactile feedback, which 
% Air typing avoids the need for external devices, but the lack of tactile feedback 
can easily lead to arm fatigue and limit input efficiency~\cite{akhoroz2024poke,yi2015atk}. 
Dudley et al. \cite{dudley2019performance} mitigated this problem by projecting a virtual keyboard onto a physical surface, which significantly improved the text entry speed by around 50\% compared to mid-air typing. 
% Input based on physical surfaces is faster than input in the air. Combining a virtual keyboard with a physical keyboard can significantly improve input efficiency. 
TouchInsight~\cite{streli2024touchinsight} further improved the typing performance on a physical surface by refining the touch area tracking accuracy, and achieved a typing speed of 37.54 WPM. \revise{
Beyond fixed surfaces, researchers explored transforming tangible props~\cite{sulaiman2008tangisoft, gil2025proptype}, arbitrary surfaces\cite{dube2022shapeshifter}, and the body~\cite{darbar2024onarmqwerty,kim2023star,lee2019quadmetric} into tactile typing interfaces.}

% uses visual tracking to enable ten-finger typing on any surface at a speed of 37.54 WPM. 

Researchers have also explored integrating physical keyboards into immersive environments. Grubert et al. \cite{grubert2018text} explored adapting standard keyboards, including physical keyboards and tablet soft keyboards, for use with an immersive head-mounted display. Their findings revealed a \revise{certain gap}
% significant drop 
in typing speed compared to typing in the real world. Researchers have explored combining with other input modalities, such as gaze and speech, to mitigate this problem. Adhikary et al. \cite{adhikary2021text} combine speech with physical keyboard input and showed a significant improvement in input speed and error rate, but voice-based input is prone to suffer from privacy problems. Kalamkar et al. \cite{kalamkar2024accented} combined gaze with a physical keyboard to improve the efficiency of special character input, but it increases the user's visual burden. 
% Combining voice with keyboard input can significantly improve input speed, but it has limitations in terms of noisy environments and privacy\cite{adhikary2021text}. 
HawKEY~\cite{pham2019hawkey} further investigated improving the flexibility of VR physical keyboard typing with a hawker’s tray to support typing while standing. 
% HawKEY\cite{pham2019hawkey} provides a tray-type virtual keyboard to support mobile scene input. 
% Voice input also has great development potential. 

Beyond the physical keyboard, researchers also investigated hands-free typing techniques, such as using eye movements. Eye-based typing shows an advantage in HMD typing, as it relies only on natural eye movements without requiring additional physical effort. However, gaze-based typing normally requires a dwell action to trigger the input event, which would reduce the input speed \cite{yu2017tap, lu2020exploration}. RingText \cite{xu2019ringtext} mitigated this problem by exploring a circular keyboard layout to improve the text entry speed with gaze without dwell. Hu et al. \cite{hu2024skimr} further explored the dwell-free eye typing technique on a virtual QWERTY keyboard in mixed reality. 
% Eye movement input frees up the hands. 
% Compared with the gaze and wait method, eye movement and sliding input are more efficient\cite{yu2017tap,lu2020exploration,xu2019ringtext}. 
% Eye-tracking input has also made progress in eyes-free input.
iText~\cite{lu2021itext} pushed the limit of eye-based typing techniques by exploring typing on an invisible keyboard. Their solution achieved a text entry speed of 13.77 WPM without requiring too much eye attention. Apart from eye-based typing, Wan et al.~\cite{wan2024exploration} recently pioneered the use of foot-based gestures for text entry in VR. Their study focused on evaluating tap-based and swipe-based text entry techniques, and the results show that bipedal tap-based input yields the best performance of 11.12 WPM in sitting posture. 
% combines eye-tracking input with an invisible keyboard, achieving a speed of 13.77 WPM. Foot input has gradually attracted attention. 
% Wan et al.\cite{wan2024exploration} implemented a foot-based text input system with an input speed of 11.12 WPM, but it can only achieve high efficiency in a sitting position. 

\sysName{} follows Wan et al.~\cite{wan2024exploration}'s concept of foot-based typing, and extends this concept from two perspectives. First, we consider that VR applications, such as gaming, often require a standing posture. Therefore, we optimized the foot-based typing experience to support both sitting and standing conditions. Second, we consider that VR applications mostly demand users' visual attention. Therefore, our explorations focus on enabling eye-free text entry in virtual scenes.

% Previous research on foot input in VR environments has faced two major challenges: first, the overall number of related studies is relatively small, and the theoretical and technical frameworks are still immature; second, existing foot input solutions are limited to a single posture. Our work aims to develop a foot-based input method for VR environments that uses similar input logic for both standing and sitting positions and achieves high input efficiency.

\subsection{Foot-based Interfaces}

We also drew inspiration from existing foot-based interfaces in the design of \sysName. 
Foot-based input had been widely explored as an alternative interaction modality, particularly in scenarios where both hands are occupied\cite{yue2023achieve,velloso2015feet}. Prior research had examined diverse approaches, including pressure-based interaction \cite{kim2018pressure,tao2012real}, heel on-floor rotation \cite{zhong2011foot}, ankle mid-air rotation \cite{scott2010sensing}, and foot tapping interaction \cite{rajanna2022presstapflick}. Yasmin et al. \cite{felberbaum2018better} further conducted a gesture elicitation study on foot-based gestures to identify the potential of foot-based gestures to support avatar control and GUI control. 

In the VR context, users' hands are often occupied by the hand-held controller or other hand-based tasks such as gesturing. Therefore, foot-based interactions were considered suitable for VR applications and had been explored in recent years' research. Christopher et al. \cite{austin2020elicitation} conducted a gesture elicitation to explore user-defined foot gestures for AR map interactions, while Shih et al.~\cite{shih2025seethroughbody} designed a floor-projected radial menu interaction with semi-transparent avatars to address occlusion. \revise{Similarly, Müller et al.\cite{muller2019mind} investigated foot-tapping input for HMDs by comparing direct floor projections with indirect floating interfaces. Their findings reveal that although direct interaction yields higher accuracy, users prefer indirect interaction due to improved comfort and reduced neck fatigue.} Recently, Chan et al.~\cite{chan2024seated} proposed Seated-WIP, a simple yet inspiring VR locomotion technique using foot stepping. Their design distinguished forefoot stepping and rearfoot stepping to support moving forward and backward with low fatigue required and high input efficiency. 

\sysName{} is largely inspired by the aforementioned literature focusing on foot-based interaction. On one hand, the interface design of \sysName{} draws inspiration from Foot Menu~\cite{zhong2011foot}, where they design a radial-based foot menu to support heel rotation interaction. On the other hand, we referred to Seated-WIP's \cite{chan2024seated} hardware solution when implementing \sysName.

\subsection{Eye-free Text Entry Techniques}

We also refer to prior eye-free text entry research when designing \sysName. Eye-free text entry refers to entering text without requiring visual attention to the keyboard or finger/cursor movement~\cite{tinwala2009eyes, xu2019tiptext, lu2017blindtype}. Literature has explored supporting eye-free text entry across various contexts, such as smartphone~\cite{zhu2018typing, ye2020qb}, tablet~\cite{li2023restype}, VR~\cite{gil2020characterizing}, and wearable devices~\cite{xu2019tiptext, xu2020bitiptext}. For example, Zhu et al.~\cite{zhu2018typing} explored supporting typing on a smartphone with an invisible keyboard, achieving an average typing of 37.8 WPM after 3 days of training. 
Li et al.~\cite{li2023restype} extend this concept to tablets and present ResType, a system that adaptively adjusts the invisible keyboard placement to facilitate eye-free text entry, with a text entry speed of 41.6 WPM. 
Gil et al.~\cite{gil2020characterizing} further extend this concept to VR and propose an invisible keyboard to support eye-free mid-air typing, with an average input speed of 41.6 WPM.
Eye-free typing has also been explored for wearable devices, as mobile application contexts often require users' attention on their surroundings. TipText~\cite{xu2019tiptext} and BiTipText~\cite{xu2020bitiptext} support eye-free thumb-to-index finger typing on a QWERTY layout keyboard, achieving 13.3 WPM and 25 WPM, respectively. \revise{Furthermore, TypeAnywhere~\cite{zhang2022typeanywhere} enables ten-finger QWERTY typing on any surface by decoding finger-tap sequences, achieving 70.6 WPM.}
Other approaches, such as BlindType~\cite{lu2017blindtype} and i'sFree~\cite{zhu2019sfree}, support eye-free remote typing by using a handheld touchpad or a smartphone as a typing proxy without requiring users' attention on it. 

The design concepts and approaches of \sysName{} are largely inspired by the aforementioned literature. In particular, we adopt a similar approach proposed by TipText~\cite{xu2019tiptext} and BiTipText~\cite{xu2020bitiptext}, where we optimize our keyboard layout by combining a user study on users’ natural ankle spatial awareness with a computer-simulated language model to reduce word ambiguity.

\section{Design Space and Research Questions}

This section elaborates on the design space of the ankle-based typing technique, including interface design and interaction design, and defines our research questions. 

\subsection{Interface Design}
\label{sec_design_space_interface}

\begin{figure}
    \centering
    \includegraphics[width=0.7\linewidth]{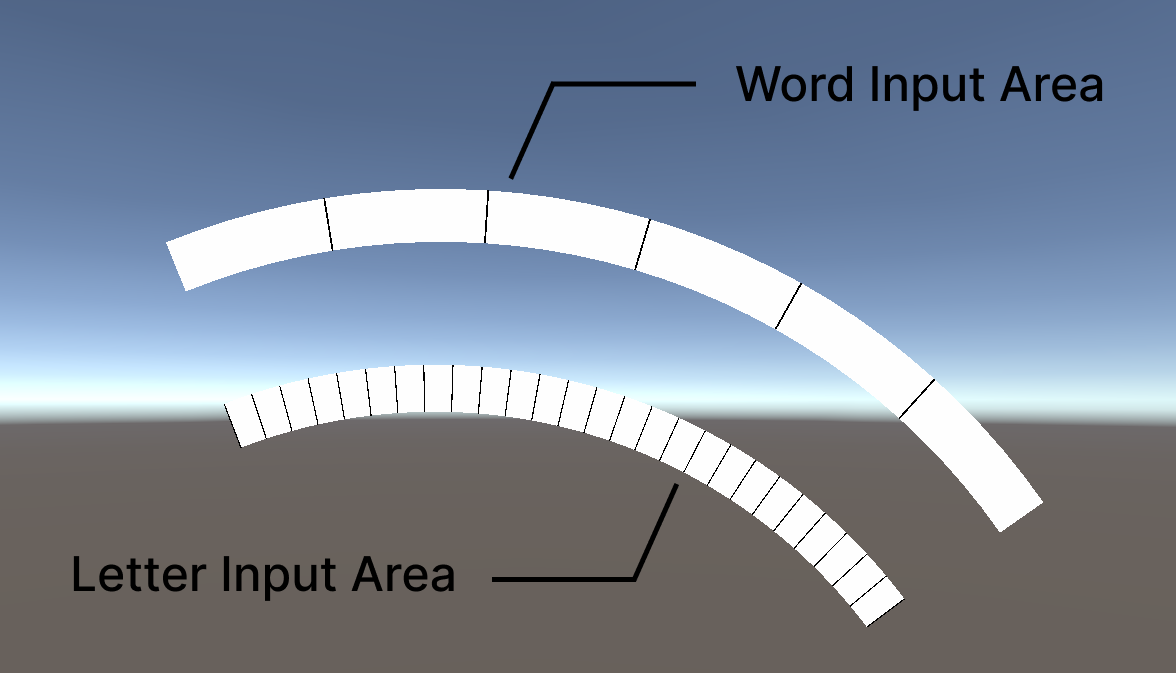}
    \Description{Illustration of our interface design concept of using two concentric radial areas for letter and word input areas separately.}
    \caption{Illustration of our interface design concept of using two concentric radial areas for letter and word input areas separately. }
    \label{fig:Layout_26Key}
\end{figure}

Our design follows a general principle of ergonomic interaction design~\cite{velloso2015interactions, shih2025seethroughbody, zhong2011foot} where we adapt a radial menu layout controlled by ankle rotation. 
In the design context of text entry interfaces, following prior works~\cite{wu2024fanpad, xu2019ringtext,watson2022feet,lopes2019feetiche}, we organize the interface into two concentric radial areas for letter and word input separately, as shown in Figure~\ref{fig:Layout_26Key}.

Importantly, our design aims to support ankle-rotation typing in both standing and sitting postures, with the cursor's position mapping to users' ankle movement. 
This raises a critical design consideration to investigate whether there is any significant difference in ankle rotation range across different postures, to determine if it is necessary to design distinct keyboard layouts for each posture. 

In this context, another design factor, keyboard layouts, was identified with considerations of balancing the pointing accuracy and the text input efficiency under eye-free conditions. Keyboard layout here refers to the number, size, and spatial arrangement of keys, as well as their corresponding letter assignments.
Specifically, layouts with larger keys may facilitate pointing speed and accuracy, while layouts with smaller keys allow alphabets to be assigned more separately across different keys, which may improve the language disambiguation and typing efficiency. Accordingly, we raise the following two research questions to guide our exploration on the interface design of \sysName:

\begin{enumerate}
[itemsep=.5em, label={\textbf{RQ\arabic*}}]
\item Are there significant differences in ankle rotation range between standing and sitting postures, and if so, how do these differences appear?

\item How to find an optimal keyboard layout considering the balance between pointing accuracy and input efficiency under the eye-free condition?
\end{enumerate}

\subsection{Interaction Design}

With the proposed interface, our goal is to investigate a user-friendly and efficient interaction design to support the text entry task in the VR context. \revise{We divide the interaction task into two sub-tasks: navigation and control, where the navigation task focuses on controlling cursor movement, and the control task focuses on command execution. For the navigation input, our goal is to provide an ergonomic input solution while minimizing users' foot movement, as suggested by prior works \cite{wan2024exploration}. Prior research on foot-based interaction revealed that heel-pivot rotation is efficient and user-friendly \cite{scott2010sensing, austin2020elicitation}, which has been widely adopted to foot-based HCI systems \cite{watson2022feet,kim2018pressure,lopes2019feetiche}. Building on this foundation,}we adopt Norman’s design principle of natural mapping~\cite{10.5555/2187809}, by explicitly mapping the cursor position to the direction in which the user’s toe points, to provide an intuitive and naturally controllable interaction experience. 

Building on it, our exploration focuses on the interaction design for control inputs, including commands for basic input tasks such as confirm and delete, and commands for switching between letter-input and word-input. In addition, we also investigate typing with unipedal and bipedal input, in both standing and sitting postures. This is motivated by prior research~\cite{wan2024exploration} suggesting that using both feet in a sitting posture would improve typing efficiency through alternating foot usage, while we hypothesize that typing with both feet in a standing posture may increase users' fatigue. To find an optimal trade-off between typing efficiency and users' preferences across postures, we formulate the following two research questions to guide our exploration on the interaction design of \sysName:

\begin{enumerate}
[itemsep=.5em, label={\textbf{RQ\arabic*}}]
\setcounter{enumi}{2}
\item What are users’ preferences for different input commands in the ankle-based text entry task, considering both unipedal and bipedal interaction? 
\item How do different input strategies (unipedal vs. bipedal) under postures (standing vs. sitting) affect users’ typing performance and their subjective preference in ankle-based text entry tasks?
\end{enumerate}

\section{Preliminary Explorations: Understanding Users' Performance and Preference of Ankle-based Foot Interaction}

Our exploration starts with two preliminary studies. One is a within-subject quantitative analysis of ankle horizontal rotation range in both standing and sitting postures across users (\textbf{RQ1}). Another one is an elicitation study to generate user-defined ankle-based foot gestures for each of the input actions (\textbf{RQ3}).

% \subsection{Participants}

% We recruited 22 participants (13 male, 9 female), aged 18-26 years(Mean = 22.00, SD =2.83), through social media. Their shoe sizes (EU) ranged from 41 to 44 (Mean = 42.00, SD=0.96) for male and from 36 to 39(Mean = 37.72, SD =1.15) for female. All participants were right-handed users. Eight participants reported prior VR experiences, and four of them had more than 0.5 years of experience across applications such as gaming, documenting, and movie watching. We run the following two preliminary studies on the same group of participants.

\subsection{Preliminary Study 1: Investigation of Ankle Horizontal Rotation Range across Posture}

\subsubsection{Participants}

We recruited 22 participants (13 male, 9 female), aged 18-26 years $(M = 22.00, SD = 2.83)$, through social media. Their shoe sizes (EU) ranged from 41 to 44 $(M = 42.00, SD = 0.96)$ for male and from 36 to 39 $(M = 37.72, SD = 1.15)$ for female. All participants were right-handed users. 

\subsubsection{Apparatus}

    To accurately track users' ankle horizontal pointing angle, we developed a customized shoe with an HTC Vive Tracker 2.0 that was mounted on the toe cap (Figure~\ref{fig:Apparatus_and_Layout}a), paired with an HTC Vive Pro 2 HMD for tracking. We built a customized program using Unity3D (2023.2.20f1c1) with SteamVR Unity plugin (version 2.8.0) on a desktop PC with an i9-14900K CPU, 64 GB RAM, and an NVIDIA GeForce RTX 4060 GPU for the experiment. An HTC Vive controller was used to trigger the data recording. The program recorded a 3-DoF rotation data of the tracker only when the controller trigger was pressed. \revise{We provided a 42 cm-high chair with fixed armrests to simulate the most common and natural sitting postures.}

\subsubsection{Study Design and Procedure}
\label{SEC:stand and seat details}
This within-subject study measures and compares 
% the ankle rotation range between standing and sitting postures. 
\revise{the range of right ankle rotation between standing and sitting postures.
Throughout the study, participants were instructed to sit/stand in a natural way with their feet naturally positioned in their most comfortable posture.
}To capture representative measurements for the range of rotation, we measured the angles at three positions on each trial: the middle-rest position, the far-left position, and the far-right position. 

Participants began the study by filling out a pre-test questionnaire with demographic information. After a brief introduction of the purpose of the study, participants were required to wear our customized tracking shoe \revise{on the right foot} and hold an HTC Vive controller. For each measurement, participants rotated their ankle to the instructed position and pressed the trigger to record the angle data. Each trial began at the middle-rest position, representing the most comfortable resting position, followed by gradually rotating to either the far-left or far-right position within the comfort range. To mitigate the order effect, the order of the left and right measurements is counter-balanced in each trial. 

Each block consisted of 10 trials, and participants were required to complete two blocks per posture (standing vs. sitting), with one-minute breaks between blocks. The posture order was counterbalanced across participants. As a result, each participant contributed 2 (blocks) × 10 (trials) × 2 (postures) = 40 trials, with each trial consisting of three recorded angles from three positions.

\subsubsection{Results}

To facilitate the data analysis, we transferred the angle data into the left range (range from far-left to middle-rest) and the right range (range from middle-rest to far-right). This yielded two dependent variables (DV), with the posture (standing vs. sitting) as the independent variable (IV). Shapiro-Wilk tests indicated except for the left range of standing, other data are normally distributed (p > .05). 

We applied Wilcoxon Signed-Rank test to compare the left range between standing and sitting. The results indicated that for the left range, standing (M = 33.66, SD = 6.52) was significantly larger than sitting (M = 30.00, SD = 5.63), Z = -3.2, p < 0.005.
For the right range, paired-samples t-test revealed the right range was greater in standing (M = 50.77, SD = 8.83) than in sitting (M = 44.58, SD = 6.91), t(21) = –4.57, p < .005.

We further look into the proportion of the left and the right angle ranges in the overall angle span. Paired-samples t-test indicates that there were no significant differences (t(21) = .31, p = .76) in left proportion for standing (M = .40, SD = .05) and sitting (M = .40, SD = .06), and no significant differences (t(17) = -.31, p = .76) in right proportion for standing (M = .60, SD = .05) and sitting (M = .60, SD = .06).

\subsubsection{Findings and Insights}

The results reveal that ankle horizontal rotation range in the sitting posture is significantly smaller than that in the standing posture, while no significant differences were observed when considering their proportions (\textbf{RQ1}). One possible reason is that participants tended to involve knee movement during standing, resulting in a larger ankle rotation range compared to sitting. This gives us several key insights.

\begin{enumerate}
[itemsep=.5em,label={\textbf{F\arabic*}}]

\item \textbf{Key Size vs. Key Number.} 
These findings suggest the need to explore separate keyboard layouts for each posture. Specifically, the differences in absolute range inspire us that the key size should be distinct for each posture to adapt to the range difference. Meanwhile, the lack of significant differences in left and right proportions across postures indicates that users' relative movements are similar across postures, suggesting the number of keys for the left and right sections can be consistent across postures to maintain a similar user experience.  

\item \textbf{Natural Movement.}
During the study, we also observed that participants preferred to rotate their ankle around the heel with their sole slightly lifted \revise{while keeping their heel as support}, rather than rotating the entire foot flat on the ground. This observation inspired us to optimize our tracking algorithm to align with users' natural movement patterns. 

\end{enumerate}

\subsection{Preliminary Study 2: User-defined Foot Gesture for Ankle-Based Typing Task}

\subsubsection{Participants}

We recruited 18 participants (12 male, 6 female) back from the previous study, aged between 18-26 years $(M = 22.67, SD =2.83)$. Their shoe sizes (EU) ranged from 41 to 44 $(M = 42.08, SD = 1.00)$ for male and from 37 to 39 $(M = 37.75, SD = 1.41)$ for female. All participants were right-handed users. Eight participants reported prior VR experiences, and four of them had more than 0.5 years of experiences, across applications such as gaming, documenting, and movie watching. 

\subsubsection{Apparatus}

We adopted a similar hardware setting to the previous study. To demonstrate the design task, we implemented a prototype radial interface with separated letter and word input areas as described in section \ref{sec_design_space_interface}. The letter area was equally divided into 26 keys, each corresponding to one letter, and we allowed the participants to control the cursor by rotating their ankle. The cursor could be switched between the letter area and the word area by pressing the space button on the keyboard.

\subsubsection{Study Design and Procedure}

The purpose of this study is to generate a series of user-preferred foot gestures for an ankle-based typing task. To this end, we adopted a gesture-elicitation study scheme~\cite{chen2021gestonhmd}, by providing the users a set of functions or action referents, and asked them to define his/her desired foot gesture using different input strategies (unipedal vs. bipedal) accordingly. We chose four referents, with their effect as shown in the Table~\ref{tab_referents}.

\begin{table}[t]
\begin{tabular}{l|p{0.7\linewidth}} 
\hline
\multicolumn{1}{c|}{\textbf{Referents}} & \multicolumn{1}{c}{\textbf{Effects}} \\ \hline
\textit{Confirm} & Enter a letter or word selected by the cursor \\
\textit{Cancel} & Delete a letter or word \\
\textit{Switch in} & Move the cursor from the letter input area to the word input area \\
\textit{Switch out} & Move the cursor from the word input area to the letter input area \\ \hline
\end{tabular}

\caption{List of referents and their effect for the foot gesture elicitation task.}
\label{tab_referents}
\end{table}

At the beginning of the session, the experiment facilitator introduced the study purpose and asked participants to complete a pre-study questionnaire collecting anonymous demographic information. Participants were then instructed to wear the customized shoe and the HMD, through which the interface was displayed in a virtual scene. To ensure that the participants fully understand the effects of the referent, we allowed them to familiarize themselves with the interface before each rating task. For those referents involving menu switching, the facilitator demonstrated the effect by pressing the space button to switch the cursor. 
To reduce the legacy bias, we adapted the \textit{Production} technique~\cite{morris2014reducing}, by requiring each participant to design 3 foot-based gestures for each referent for both unipedal and bipedal input, and rate them according to their preference. Referents were presented in the same order to each participant.
The study was conducted without posture constraints. This setting allowed participants to generate a broader range of gesture ideas, which we later evaluated and analyzed in the context of both sitting and standing use cases. After they completed each of the referents, we performed a semi-structured post-task interview to elicit feedback about their experience, including their design consideration and qualitative feedback on the design.

\subsubsection{Results and Findings}

\begin{figure*}
    \centering
    \includegraphics[width=\linewidth]{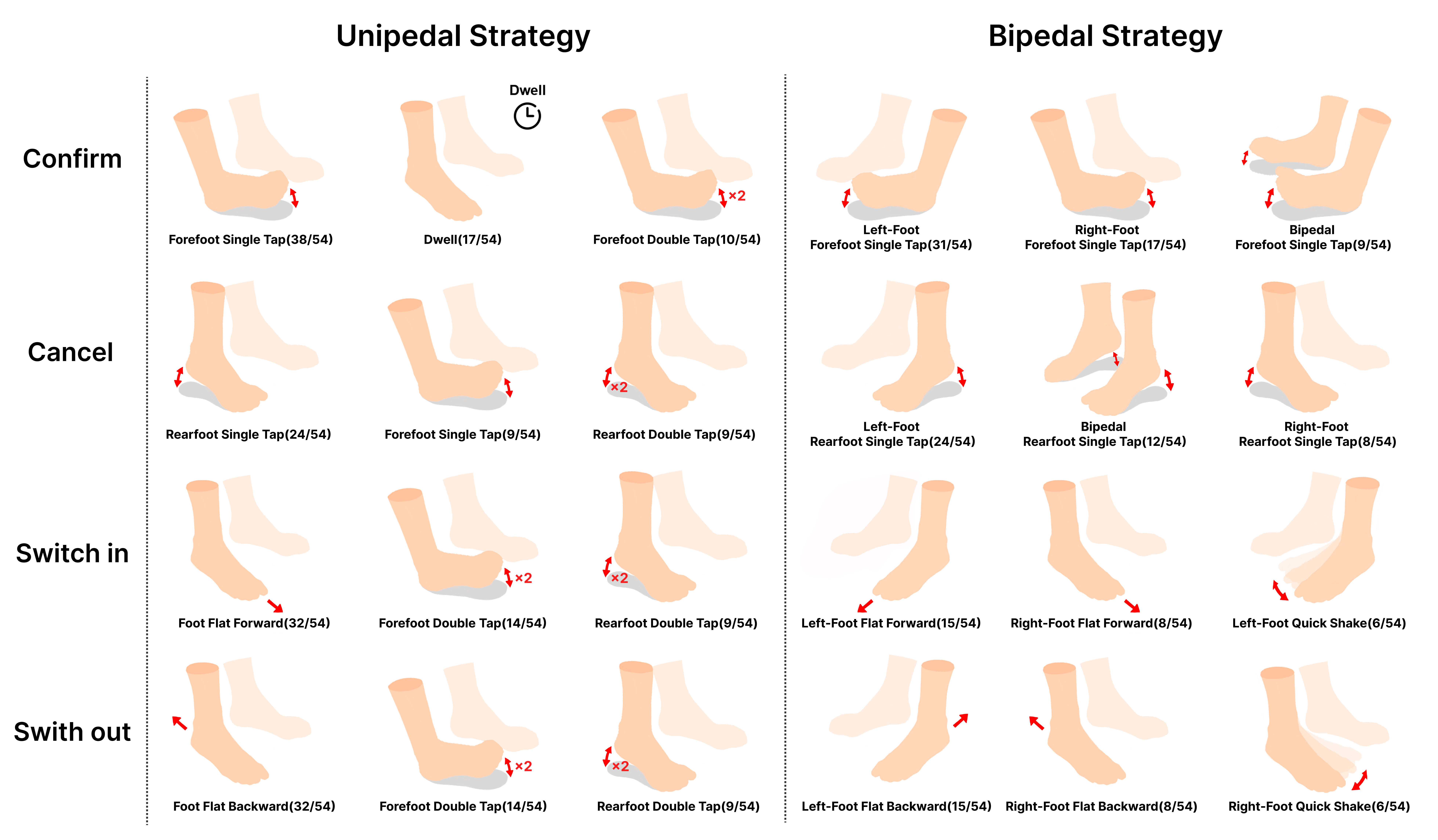}
    \Description{A chart illustrating the top three ranked gestures for four commands (Confirm, Cancel, Switch in, Switch out) under Unipedal and Bipedal strategies. 
In Unipedal mode, the highest-rated gestures are Forefoot single tap for Confirm (score 38), Rearfoot single tap for Cancel (24), and Foot flat movements for Switching (32). 
In Bipedal mode, the Left foot dominates, with Left-foot forefoot tap for Confirm (31) and Left-foot rearfoot tap for Cancel (24). 
Full scoring breakdown for the top 3 candidates is displayed for each category.}
    \caption{Foot gesture elicitation study results. This figure shows the top 3 gesture illustrations for each referent under both unipedal and bipedal strategies with their weighted preference score. Each gesture had a maximum score of 54.}
    \label{fig:FootGesture}
\end{figure*}

We calculated a weighted score of each gesture based on participants' preference, where we assigned their first preference with score of 3, their second preference with score of 2, and their least preferred with score of 1. We then select the top 3 gesture candidates for each referent under both unipedal and bipedal strategies, as illustrated in Figure~\ref{fig:FootGesture}. These results and the post-task interview revealed several key insights for \textbf{RQ3}.

\begin{enumerate}
[itemsep=.5em,label={\textbf{F\arabic*}}]
\setcounter{enumi}{2}
\item \textbf{Symmetricity.} 
Participants consistently tended to produce symmetrical gestures for paired referents, regardless of whether the input was unipedal or bipedal. For example, for paired referents such as \textit{Confirm} and \textit{Cancel}, users tend to use symmetric gestures \textit{forefoot single tap} and \textit{rearfoot single tap} for both unipedal and bipedal input. For the other paired referents, \textit{Switch in} and \textit{Switch out}, the gestures with the highest score are \textit{foot flat forward} and \textit{foot flat backward} for both input strategies, which are also symmetric gestures. This insight also aligns with the results from other gesture elicitation studies that yield gestures with symmetricity~\cite{wobbrock2009user, chen2021gestonhmd}.

\item \textbf{Natural Mapping.} 
We also observed that participants prefer gestures with intuitive spatial mappings. Since the word-input area is positioned above the letter-input area, participants have a consistent preference to push forward their feet to enter the word area and backward to return to letters, regardless of whether the input is unipedal or bipedal. This finding also aligns with  Norman’s design principle of natural mapping~\cite{10.5555/2187809}.

\item \textbf{Foot Role Balancing.} 
With bipedal input, participants tended to use their dominant foot to perform precise input, such as cursor control, while assigning less precise input, such as control commands, to the non-dominant foot, thereby balancing interaction workload. While participants generally agree that bipedal input would improve input efficiency, the interview revealed that 11 participants prefer unipedal input in standing conditions, as the other foot was needed for physical support, whereas 4 participants mentioned that they prefer bipedal input for both standing and sitting postures to maintain design consistency and reduce learning fatigue.

\end{enumerate}

\section{Study 1: Keyboard Layout Optimization}

To design an optimal keyboard layout for \sysName{} that balances the pointing efficiency and input efficiency under eyes-free conditions, we adopted a data collection study to understand users' natural expectation of ankle-rotation-based input through their spatial locatability on the keys of an ankle-based keyboard without visual feedback.

\subsection{Participants}

We recruited 12 participants (8 male, 4 female), aged between 19-25 years $(M = 20.92, SD = 2.31)$, through social media. Their shoe sizes (EU) ranged from 41 to 44 $(M = 42.00, SD = 0.96)$ for male and from 36 to 39 $(M = 37.72, SD = 1.15)$ for female. All participants were right-handed users. Three of them reported prior VR experiences. 

\begin{figure*}[t]
    \centering
    \includegraphics[width=1\linewidth]{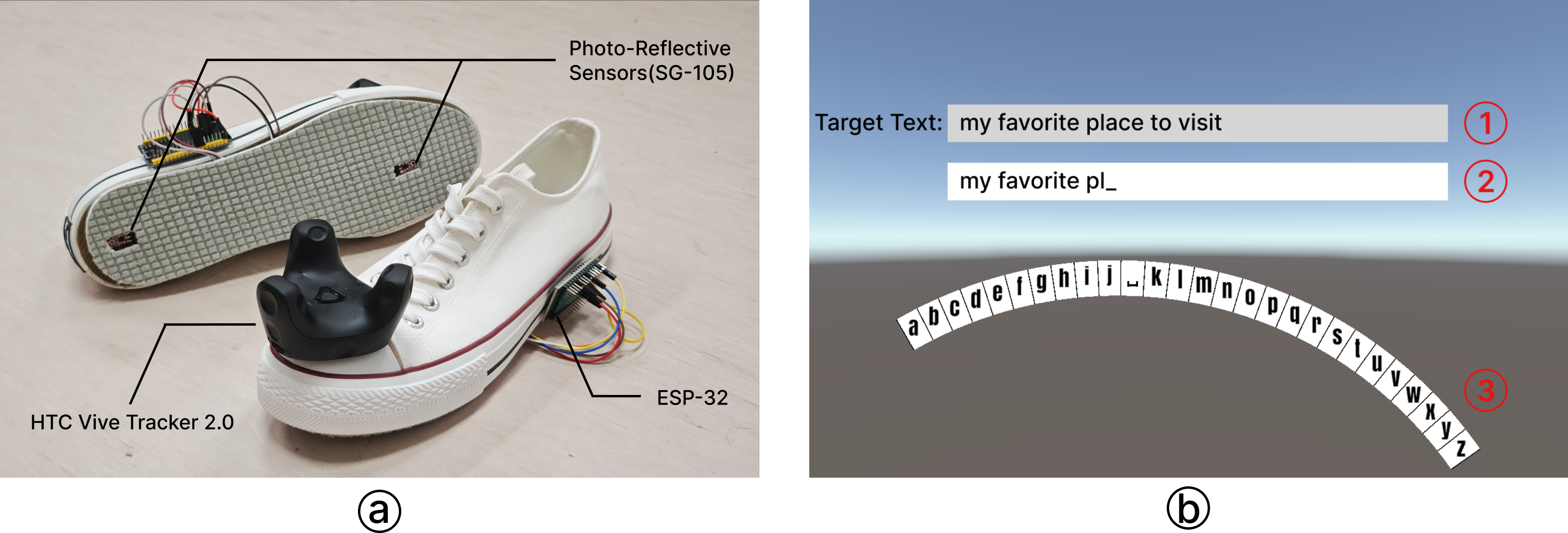}
    \Description{a.The shoe prototype for AnkleType, including two photo-reflective sensors at the fore and rear of the shoe, an ESP-32 to process the sensor signal, and an HTC Vive Tracker 2.0 to enable ankle rotation tracking. This implementation was based on Chan et al.'s [7] solution. b.The interface for the keyboard layout optimization study. ①The target text to be transcribed. ②The current input text. ③The initial keyboard layout, where we evenly distributed 27-key (26 letters + space bar) across the keyboard area.}
    \caption{\textcircled{a}The shoe prototype for \sysName, including two photo-reflective sensors at the fore and rear of the shoe, an ESP-32 to process the sensor signal, and an HTC Vive Tracker 2.0 to enable ankle rotation tracking. This implementation was based on Chan et al.'s~\cite{chan2024seated} solution.\textcircled{b}The interface for the keyboard layout optimization study.  \textcircled{1}The target text to be transcribed, \textcircled{2}The current input text, \textcircled{3}The initial keyboard layout, where we evenly distributed 27-key (26 letters + space bar) across the keyboard area. }
    \label{fig:Apparatus_and_Layout}
\end{figure*}

\subsection{Apparatus}
\label{SEC:layout_apparatus}

We implemented a customized shoe with an HTC Vive Tracker 2.0 that was mounted on the toe cap to enable rotation tracking. To detect the two foot gestures for inputting letter and deleting letter with the highest user preference, \textit{forefoot single tap} and \textit{rearfoot single tap}, we used two infrared photo-reflective sensors (SG-105) embodied at the back and front of the shoe’s sole (Figure~\ref{fig:Apparatus_and_Layout}\textcircled{a}), similar to Chan et al.~\cite{chan2024seated}. An ESP-32 chip was used to process and transfer the sensor data to the desktop interface through the serial port. We built the experiment interface using Unity3D (2023.2.20f1c1) with SteamVR Unity plugin (version 2.8.0) on a desktop PC with an i9-14900K CPU, 64 GB RAM, and an Nvidia GeForce RTX 4060 GPU for the experiment. The implementation of the rotation tracking algorithm follows the inspiration of \textbf{F2}.

% \begin{figure}
%     \centering
%     \includegraphics[width=0.7\linewidth]{figs/Apparatus.png}
%     \caption{The shoe prototype for \sysName, including two photo-reflective sensors at the fore and rear of the shoe, an ESP-32 to process the sensor signal, and an HTC Vive Tracker 2.0 to enable ankle rotation tracking. This implementation was based on Chan et al.'s~\cite{chan2024seated} solution. }
%     \label{fig:Apparatus}
% \end{figure}

% \begin{figure}
%     \centering
%     \includegraphics[width=0.7\linewidth]{figs/Layout_Blind_Exp1.png}
%     \caption{The interface for the keyboard layout optimization study.  \textcircled{1}The target text to be transcribed, \textcircled{2}The current input text, \textcircled{3}The initial keyboard layout, where we evenly distributed 27-key (26 letters + space bar) across the keyboard area.}
%     \label{fig:Layout_Blind_Exp1}
% \end{figure}

\subsection{Initial Keyboard Layout Design}

% \subsubsection{Layout Design}

As suggested by prior research on data-driven keyboard layout design~\cite{xu2019tiptext}, we began our exploration with a full alphabet keyboard. For the initial layout, we distributed 27 keys (26 letters + space bar) evenly across the keyboard area for both standing and sitting postures (Figure~\ref{fig:Apparatus_and_Layout}\textcircled{b}). \textbf{F1} revealed that the segmentation between the left and right sections of the keyboard followed the same proportion across postures. This would result in an equal number of keys on each side across postures. We assign the space bar to the middle-rest position as we consider that word input naturally ends with a space for segmentation, and the middle-rest is the most intuitive position for this input. This setting also ensures that each word entry begins from the middle-rest position, offering a consistent spatial reference for subsequent key inputs. Notably, considering the movement range difference across participants and postures, the absolute size of the keyboard was proportionally scaled to fit before each trial.

% \subsubsection{Tasks}

\subsection{Task and Procedure}

The purpose of this study is to examine users' spatial locatability on the keys of the ankle-based keyboard without visual feedback. To this end, we designed an eyes-free (i.e., no cursor was visible) within-subject text entry task using a Wizard of Oz keyboard~\cite{xu2019tiptext, zhu2018typing}. Participants were required to transcribe 2 blocks of 10 phrases in both standing and sitting postures respectively, for a total of 40 phrases. Of these, 32 phrases were picked randomly from MacKenzie's phrase set~\cite{mackenzie2003phrase}, while the rest 8 phrases, as suggested by previous works~\cite{xu2020bitiptext, goel2013contexttype, goel2012walktype}, were randomly selected pangrams\textbf{\footnote{\url{https://mseffie.com/assignments/calligraphy/Plethora\%20of\%20Pangrams.pdf}}} to ensure that every letter had a minimum presence of 15 times each. \revise{All selected phrases were case-insensitive and contained no numbers or symbols.}The order of standing and sitting conditions was counterbalanced using a Latin Square to mitigate order effects. \revise{The requirements for standing and sitting postures are the same as in previous experiments(Section~\ref{SEC:stand and seat details}). }

 Prior to the study, participants completed a demographic questionnaire. After a brief introduction, they were required to wear our customized shoes and an HTC Vive HMD for calibration. During calibration for each trial, we recorded each participant’s far-left, middle-rest, and far-right positions within their comfort range, then proportionally scaled the size of the keyboard to match their individual range. Participants then completed five phrases for training, during which they could see the keyboard layout and a cursor highlighting the currently selected key. Inspired by our exploration of \textbf{RQ3}, participants were instructed to navigate the cursor by rotating their ankle and confirmed the input with a \textit{forefoot single tap} gesture. During training, incorrect letters were displayed in red color to inform the user, but we did not provide a delete function. 

In the formal study, participants were provided a static reference keyboard layout without a visual cursor to indicate the currently selected key. They were asked to select the letter by rotating their ankle and perform a \textit{forefoot single tap} gesture on an imaginary key location for selection based on their natural spatial awareness. The system always displayed the correct letters regardless of their actual tap location. Different from the training process, we did not highlight incorrect letters in the formal study. Participants were instructed to keep typing regardless of whether they felt the inputted key was correct or not. They were required to complete the whole letter sequence for every word, without word auto-completion. Therefore, we only displayed the letter area on the interface for this study. After each phrase, the system switched to the next phrase until all phrases were completed. Participants were allowed a 3-minute break between blocks. The study lasted approximately 80 minutes, and participants were compensated with 15 USD for their time.

\subsection{Results}

\begin{figure*}[t] % [h!] is a placement specifier (here, try to place it here)
    \centering % Centers the entire figure
    \begin{subfigure}[b]{\textwidth} % [b] for bottom alignment, 0.45\textwidth sets width
        \centering
        \includegraphics[width=\linewidth]{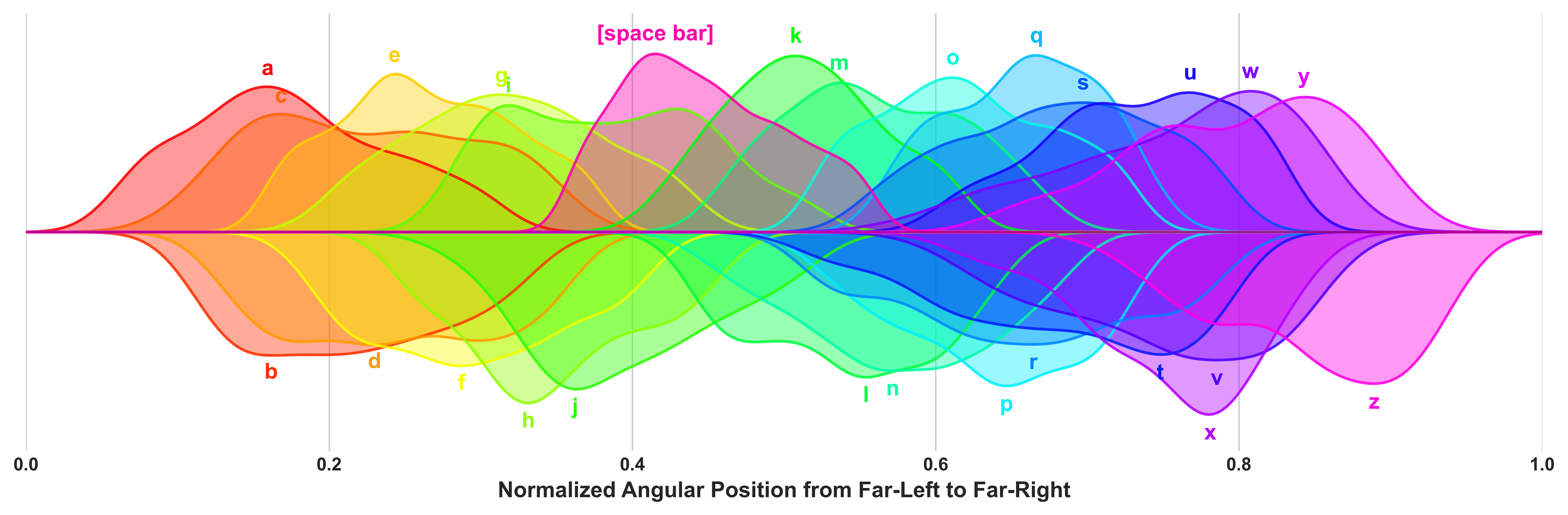} % Include your image
        \caption{Distribution of foot tapping positions for standing condition.}
        \label{fig:distribution_stand}
    \end{subfigure}
    \hfill % Adds horizontal space between subfigures on the same line
    \begin{subfigure}[b]{\textwidth}
        \centering
        \includegraphics[width=\linewidth]{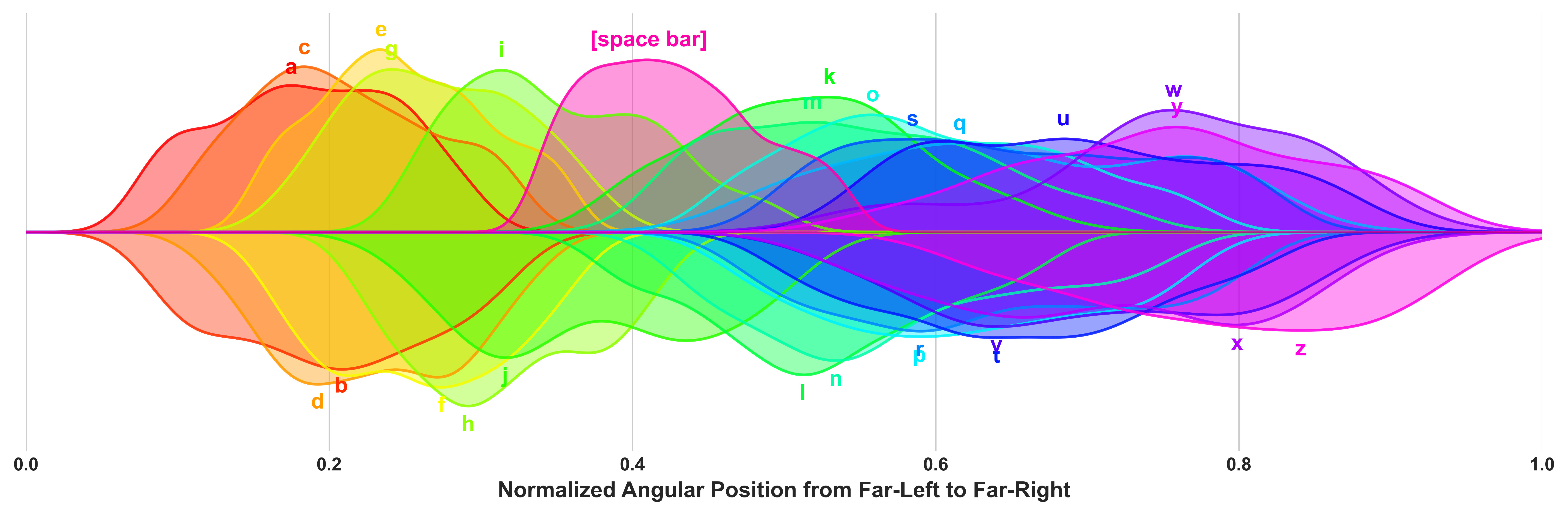}
        \caption{Distribution of foot tapping positions for sitting condition. }
        \label{fig:distribution_sit}
    \end{subfigure}
    \Description{Distribution of foot tapping positions with 95\% confidence interval in a 26-key Alphabetical-order keyboard. (a) Distribution of foot tapping positions for standing condition. (b) Distribution of foot tapping positions for sitting condition.}
    \caption{Distribution of foot tapping positions with 95\% confidence interval in a 26-key Alphabetical-order keyboard. (a) Distribution of foot tapping positions for standing condition. (b) Distribution of foot tapping positions for sitting condition. }
    \label{fig:distribution}
\end{figure*}

We collected 18,582 foot tapping positions from the study, which were distributed along a 1D space and varied across participants. To facilitate the comparison, we align the data points from different users by applying min-max normalization across users to form a general distribution for standing and sitting posture, respectively, as shown in Figure~\ref{fig:distribution}. For clarity, we present each key’s distribution with a 95\% confidence interval, and the adjacent keys are shown on different sides with distinct colors. The result shows that the tapping distributions vary across standing and sitting postures. The tapping locations between letters are noisy with significant overlaps among neighboring distributions. This finding also highlights the need to answer \textbf{RQ2}, as typing with 27 keys is not feasible, particularly in an eye-free condition. 

Despite the noise, the distributions largely followed the alphabetical order, with some keys being indistinguishable from each other and some of these characteristics consistent between standing and sitting conditions. For instance, the distributions of “p” and “r” almost completely overlapped in each condition. This highlights the necessity of incorporating a language model to decompose the input into keys with letter ambiguity. Moreover, we particularly observed that apart from the space bar, the key with the smallest standard deviation in the standing posture was the letter ``q'' ($\sigma = .070$), while in the sitting posture it was the letter ``e'' ($\sigma = .075$). In contrast, the key with the largest standard deviation in the standing posture was the letter ``r'' ($\sigma = .119$), and in the sitting posture it was the letter ``w'' ($\sigma = .169$). 
% Note that letter ``e'' and ``f'' are adjacent letter, which was place next to the middle-rest position, indicating the good spatial reachability of ankle rotation in this position for each posture. Although ``y'' and ``w'' are not adjacent, they are close to each other which are both place near the far-right position, indicating a general poor spatial reachability in this position. 
These findings suggest the reachability varies at different positions, highlighting the need to consider the spatial reachability separately for each posture condition. We later integrated this spatial information with a language model to determine the final layout.

% As the same time, key positional information should be considered to design a layout that balances spatial reachability and language model efficiency.

\subsection{Determine \sysName{} Keyboard Layout}

There exists a trade-off between keyboard layout and typing efficiency. Layouts with larger keys facilitate pointing speed and accuracy, but reduce the total number of keys, thereby lowering alphabet separability and increasing language ambiguity, which eventually slows input speed. In contrast, layouts with smaller keys allow letters to be distributed more separately across the different keys. This will improve language disambiguation and typing efficiency but reduce key size, thereby reducing pointing speed and accuracy. Based on these insights, we explore an optimal keyboard layout for our task by jointly considering the spatial model in our ankle-based typing context and the language model. 

Before deriving the two models, we first determined the acceptance range of the key number in our context. As mentioned above, the most precise key has a minimum standard deviation of around 0.07 for both standing and sitting positions. We defined the minimum spatial resolution as the condition where the overlap between two adjacent keys does not exceed half of their size. Based on this criterion, we set the maximum number of keys to 14. Moreover, as suggested by previous works~\cite{xu2019tiptext},  those keyboards with letter key numbers lower than 5 would suffer from high input ambiguity. Therefore, we set our minimum acceptance letter key number to 5. Note that in our case, the space bar takes up one key position. As a result, we set our acceptance range of key numbers from 6 to 14 in our context. 

\subsubsection{Word Disambiguation Score}

We sequentially mapped the 26 letters onto keyboards with the number of keys ranging from 5 to 13 (reserve one key for the space bar), resulting in 16774590 possible letter layout combinations. However, it is impossible to run a user study to test through these options. Instead, we adapted a simulation-based approach, as suggested by previous works~\cite{xu2019tiptext, qin2018optimal}, using a computer-simulated typing test to compare the theoretical performance across all different layouts. Specifically, we select the top 10,000 words from the American National Corpus~\cite{anc} and simulate the key sequence required to input each of them on every candidate letter layout. For each sequence, the system generated a list of candidate words (due to input ambiguation) that matched the input and were ordered by word frequency. We then recorded whether the intended word appeared within the top three entries of this list. The proportion of successful occurrences was computed as the word disambiguation score $L^k$ for each letter layout, where $k$ is the number of keys. Finally, we select the top 100 $L^k$ in each $k$ as letter layout candidates, 900 in total, denoted as $J$.

\subsubsection{Integrate with Spatial Information}

The exploration of spatial information starts by grouping the tapping positions whose spatial distributions are concentrated. To this end, we applied an unsupervised Gaussian Mixture Model (GMM) clustering approach, and iteratively set the number of clusters from 6 to 14 for both standing and sitting postures. This process yielded 18 cluster layout candidates, with 9 for each posture, denoted as $I$.

For each cluster layout candidate in each posture, we first determine the cluster with the largest number of points as the space key, and we exclude the space key for the following calculation. Then, we sequentially mapped the letters from all letter layout candidates with the same key number setting on it. This process resulted in 1,800 combinations in total, with 900 for each posture. For each cluster-layout and letter-layout pair $(i, j|i \in I, j \in J)$, we computed a spatial matching score $S(i, j)$ that integrates tapping accuracy with corpus frequency. Specifically, for each letter $l$ in $(i, j|i \in I, j \in J)$, we calculated the weighted letter score $s_l(i, j)$ as:

\begin{equation}
s_l(i, j) = \frac{n_l(i, j)}{N_l} \times f_l
\end{equation}

\noindent where $n_{l}$ is the number of tapping samples for letter $l$ that fall within the assigned key cluster, $N_{l}$ is the total number of tapping samples for $l$, and $f_{l}$ is the frequency of $l$ in the corpus\footnote{https://pi.math.cornell.edu/~mec/2003-2004/cryptography/subs/frequencies.html}. We then summarize the score across all letters to compute the cluster-layout and letter-layout pair$(i, j|i \in I, j \in J)$ spatial matching score $S(i, j)$:

\begin{equation}
S(i, j) = \sum_{l \in A} s_l(i, j)
\end{equation}

\noindent where $A$ denotes the set of all 26 letters. 

With the inspiration from \textbf{F1} and the consideration of following design consistency, we decided to keep the number of keys and letter layout consistent for standing and sitting posture. Therefore, we jointly consider $S(i, j)$ across postures by accumulating those that have the same number of key $k$, denoted as $S_{joint}^{k}(i, j)$. 

\begin{equation}
S_{joint}^k(i, j) = S_{stand}^k(i, j) + S_{sit}^k(i, j)
\end{equation}

For each $(i, j|i \in I, j \in J)$ under each $k$, we calculate their sum of its word disambiguation score $L^k(j)$ and $S_{joint}^k(i, j)$ as the final score:

\begin{equation}
F^k(i, j) = L^k(j) + S_{joint}^k(i, j)
\end{equation}

\begin{figure}
    \centering
    \includegraphics[width=1\linewidth]{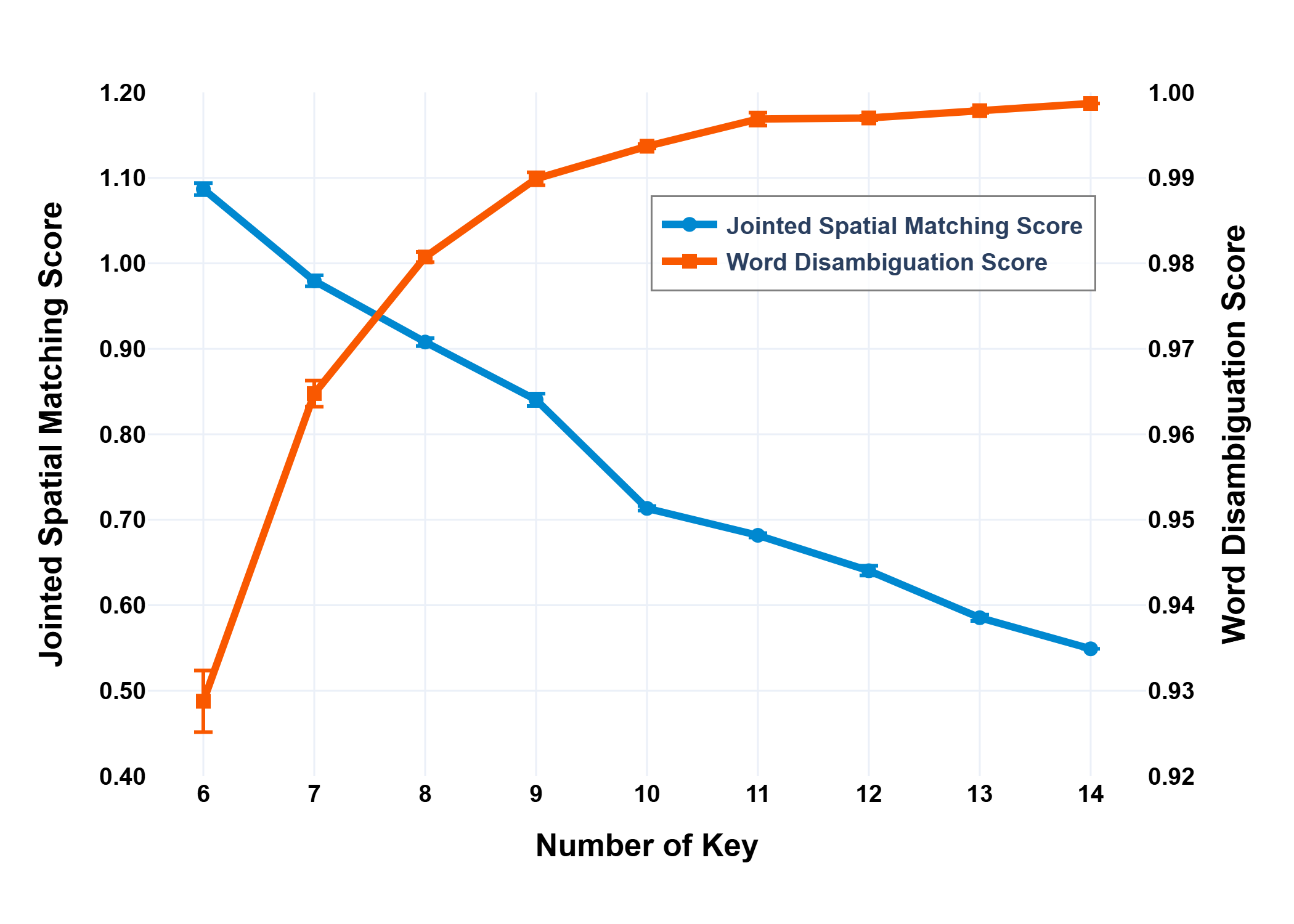}
    \Description{Word disambiguation score $L^k$ and jointed spatial matching score $S_{joint}^k$ of each key number. The figure shows two lines for the change of jointed spatial matching score and word disambiguation score. The x-axis of the chart is the number of key, ranged from 6 to 14. The y-axis indicates the score value. The slope of the jointed spatial matching score has a sharp drop from number 9 to 10. The upward trend of the word disambiguation score gradually slows down when the number of keys is larger than 9. Note that we did not include the space key when calculating these metrics. Therefore, the actual number of keys for calculating these metrics should be minus one.}
    \caption{Word disambiguation score $L^k$ and jointed spatial matching score $S_{joint}^k$ of each key number. Note that we did not include the space key when calculating these metrics. Therefore, the actual number of keys for calculating these metrics should be minus one.}
    \label{fig:language-layout}
\end{figure}

\begin{figure}[htb]
    \begin{subfigure}[b]{0.45\textwidth}
        \includegraphics[width=\textwidth]{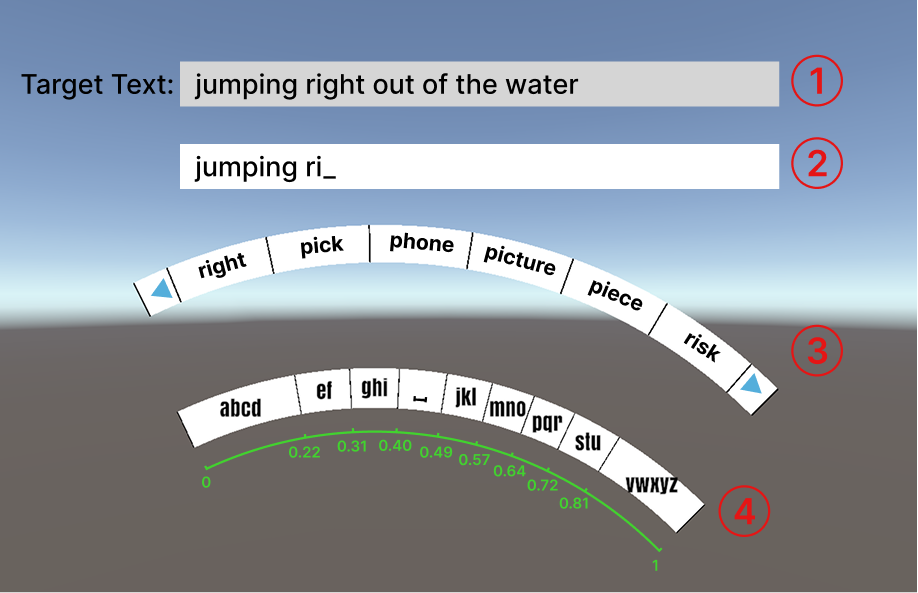}
        \caption{\sysName{} layout of standing.}
        \label{fig:Standing_Final}
    \end{subfigure}
    \hfill 
    \begin{subfigure}[b]{0.45\textwidth}
        \includegraphics[width=\textwidth]{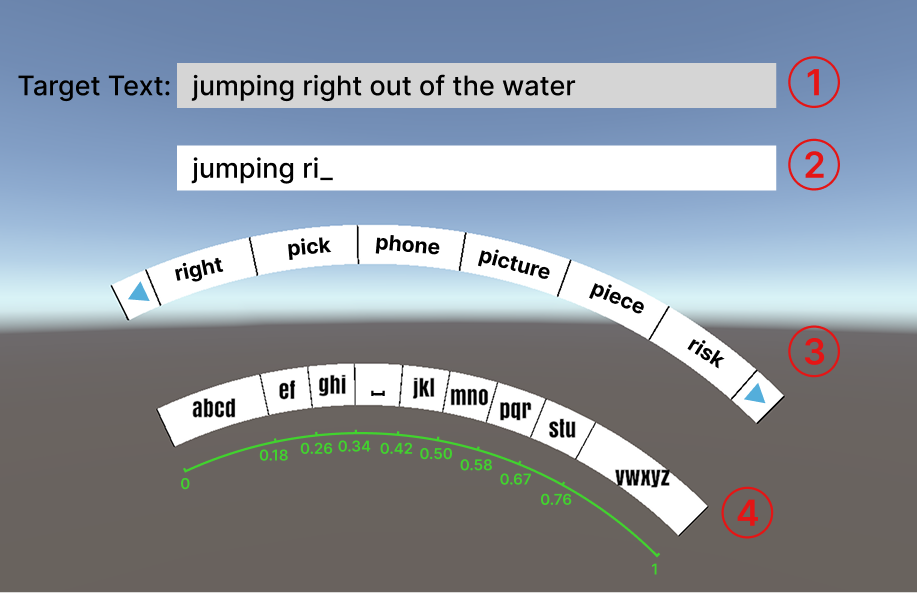}
        \caption{\sysName{} layout of sitting.}
        \label{fig:Sitting_Final}
    \end{subfigure}
    \Description{Illustration of the user study interface with the optimized keyboard layout for both standing (a) and sitting (b) conditions. ①The target text to be transcribed. ② The current input text. ③ The word input area. Two side-buttons at left and right are for page navigation. ④The optimized letter input area. The green scale below indicates the normalized angular position from far-left to far-right}
    \caption{Illustration of the user study interface with the optimized keyboard layout for both standing (a) and sitting (b) conditions. \textcircled{1} The target text to be transcribed, \textcircled{2} The current input text, \textcircled{3} The word input area. Two side buttons on the left and right are for page navigation, \textcircled{4}The optimized letter input area. The green scale below indicates the normalized angular position from far-left to far-right. }
    \label{fig:Final_Layout}
\end{figure}

Then for each $k$, we average the top 10 $F^k(i, j)$ and plot in Figure ~\ref{fig:language-layout}. As shown in the plot, the word disambiguation score increases with the number of keys, whereas the spatial score decreases. Notably, we observed a dramatic change in the spatial score at the key number of 9, where the growing trend of the word disambiguation score starts to slow down. Based on this trade-off, we selected 9 as the final key number. We then looked into the top 10 letter layouts for the key number of 9, and we were pleasantly surprised to find that one of them had a top 1 score in $L^9$ (rank 6 in $F^9$). As a result, we chose this cluster-layout and letter-layout pair as the final layout (\textbf{RQ2}) (see Supplementary Materials for other pair candidates). 
% \hl{filtered those layouts that appeared in both standing and sitting positions,} and chose the one with the highest score as the final layout (\textbf{RQ2}). 
Please refer to Figure~\ref{fig:Final_Layout} for the detailed keyboard layout.

\section{Study 2: Exploring Ankle-Based Text Entry Technique when Standing and Sitting}

With the optimal keyboard layout, we conducted a within-subject comparison study to evaluate the performance and user experience of different input strategies (unipedal vs. bipedal) under standing and sitting postures.  

\subsection{User-elicited Ankle-Based Text Entry Technique}

Based on the user-defined gesture set (Figure~\ref{fig:FootGesture}), we first designed two ankle-based text entry techniques for both unipedal and bipedal using the most popular gestures. The same techniques were used across postures, yielding four ankle-based text entry techniques for this comparison study. 

\textbf{Unipedal Stand/Sit (UPStand/UPSit).} Users navigate the cursor by rotating their right ankle. A right-foot \textit{forefoot single tap} gesture for entering a letter or word, while a right-foot \textit{rearfoot single tap} gesture for deleting. Users perform a \textit{foot flat forward} gesture with their right foot to enter the word input interface, and a right-foot \textit{foot flat backward} gesture to return to the letter input interface.

\textbf{Bipedal Stand/Sit (BPStand/BPSit).} Users navigate the cursor by rotating their right ankle. A left-foot \textit{forefoot single tap} gesture for entering a letter or word, while a left-foot \textit{rearfoot single tap} gesture for deleting. Users perform a \textit{foot flat forward} gesture with their left foot to enter the word input interface, and a left-foot \textit{foot flat backward} gesture to return to the letter input interface.

\subsection{Participants}

We recruited 16 participants (10 male, 6 female) through word of mouth, aged from 19-25 years (M = 22.13, SD = 2.23). Their shoe sizes (EU) ranged from 40 to 43 (M =
41.70, SD = 1.16) for male and from 37 to 39 (M = 37.83, SD = 2.23) for female. Their mean self-assessed typing proficiency is 6.5/10. All participants were right-handed users. One of them explicitly mentioned being familiar with VR text entry using controller ray casting. The other participants had no experience with VR text entry. 

\subsection{Apparatus}

We adopted similar hardware and software settings as our previous study (Section~\ref{SEC:layout_apparatus}). To simulate a real-world typing experience, we implemented an interface for this study with a language statistical decoding algorithm, based on a Bayesian maximum likelihood estimation~\cite{goodman2002language}, similar to~\cite{wan2024exploration}. The lexicon we used was the top 10k words from the American National Corpus~\cite{anc}. With the decoding algorithm, the word-input area displays the predicted words ordered from left to right by their probabilities. The GUIs for this study are shown in Figure~\ref{fig:Final_Layout}.

\subsection{Study Design and Procedure}

We conducted a within-subjects study with posture (standing vs. sitting) and strategies (unipedal vs. bipedal) as two independent variables, yielding a 2 $\times$ 2 design with 4 conditions in total. \revise{The requirements for standing and sitting postures are the same as in previous experiments(Section~\ref{SEC:stand and seat details}).}The order of these conditions was counterbalanced using a Latin-Square design to mitigate the order effects. In each condition, participants were asked to transcribe 10 phrases randomly selected from MacKenzie's phrase set~\cite{mackenzie2003phrase}. \revise{All selected phrases were case-insensitive and contained no numbers or symbols.} Since the goal of this study was to explore input strategies across postures rather than measuring the eye-free performance, participants were provided with visual feedback of the keyboard layout during the task.

Prior to the study, participants completed a demographic questionnaire. After a brief introduction, they were required to wear our customized shoes and an HTC Vive HMD for calibration. During the calibration, we measured each participant’s right-ankle rotation range in both sitting and standing conditions. This calibration data was used to proportionally scale the keyboard layout to fit participants' individual movement range for the corresponding condition. Before each condition, participants completed a short training session consisting of 5 practice phrases to familiarize themselves with the input method.

During the main study, participants were instructed to enter phrases as quickly and accurately as possible. After each condition, they were asked to complete the post-study questionnaires, including a NASA-TLS questionnaire and a System Usability Scale (SUS) questionnaire. A two-minute break was provided between conditions. After finishing four conditions, we conducted a semi-structured interview to collect their subjective feedback. The study lasted approximately 75 minutes, and participants were compensated with 15 USD for their time.

\subsection{Results}

\begin{figure*}
    \centering
    \includegraphics[width=\linewidth]{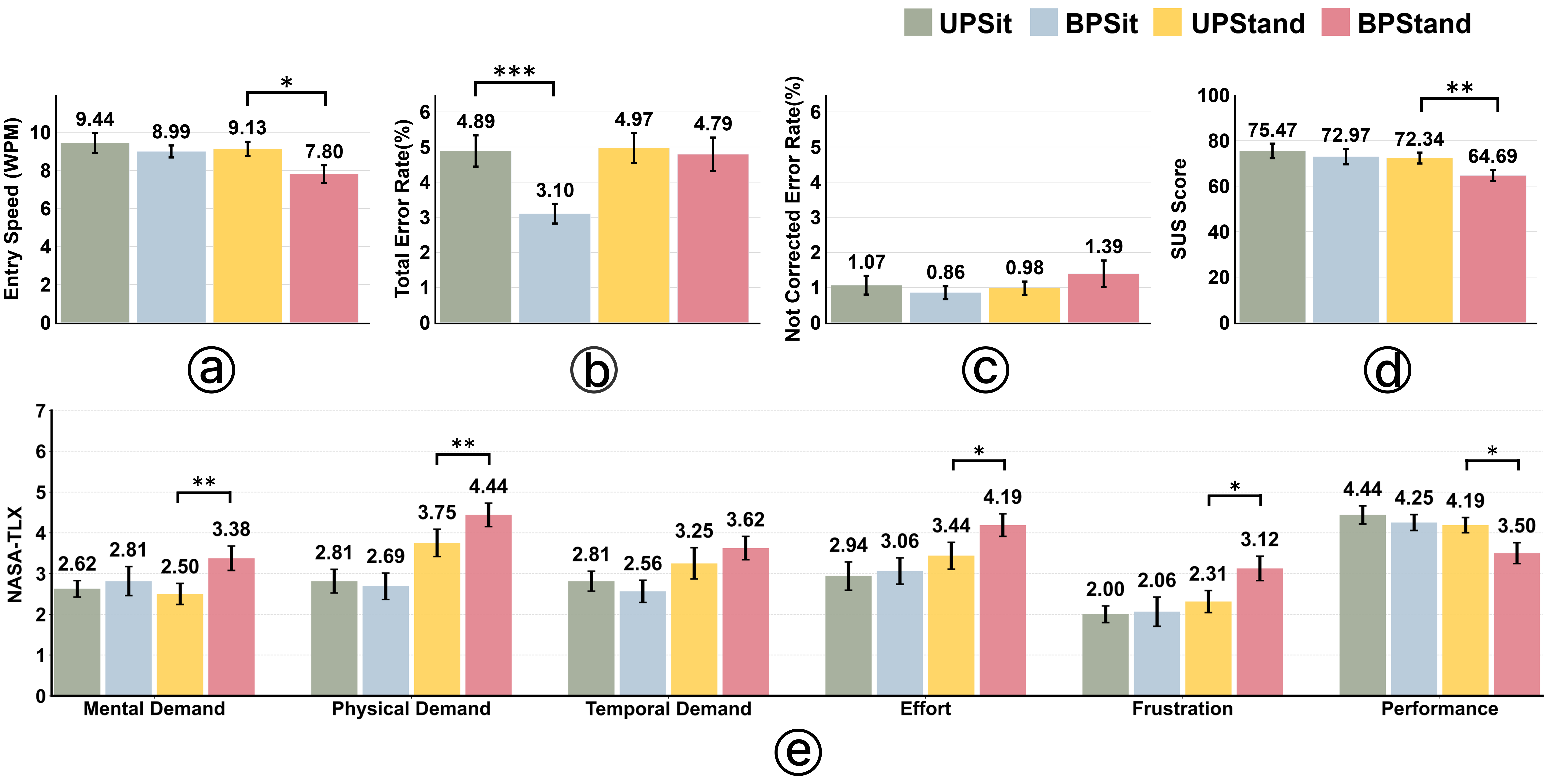}
    \Description{Bar charts comparing Entry Speed, Error Rates, SUS scores, and NASA-TLX ratings across four conditions: Unipedal/Bipedal Sitting and Unipedal/Bipedal Standing. }
    \caption{The means of \textcircled{a} Entry speed, \textcircled{b} TER, \textcircled{c} NCER, \textcircled{d} SUS Score and \textcircled{e} NASA-TLX scores. ***, **, and * represent a .001, .01, and .05 significance level, respectively.}
    \label{fig:Study2_Result.png}
\end{figure*}

We measure the entry speed in words per minute (WPM), the error rate in total error rate (TER) and not corrected error rate (NCER), the workload in NASA-TLX score, and usability in SUS score. A Shapiro-Wilk test was performed before the analysis, indicating the WPM, TER, and SUS score were normally distributed (p > .05), while the NCER was not normally distributed (p < .05). The results are shown in Figure~\ref{fig:Study2_Result.png}.

\subsubsection{Entry Speed}

We first applied an RM-ANOVA test on the WPM to see the effect of posture and strategies on the entry speed. The results show that posture does not significantly affect the entry speed $(F(1, 15) = 4.44, p = .052, \eta_p^2 = .23)$, while strategies have a significant effect on WPM ($F(1, 15) = 8.86, p < .05, \eta_p^2 = .37$). A post-hoc pairwise comparison indicated that the entry speed of unipedal ($M = 9.28, SD = .41$) is significantly higher ($p < .05$) than that of bipedal ($M = 8.40, SD = .33$). 

% overall unipedal faster than bipedal, but less accurate than bipedal

% standing: unipeda faster
% sitting: no speed difference

We further conducted a comparison across strategies for each posture. Paired-samples t-test indicate the WPM of unipedal ($M = 9.13, SD = 1.50$) is significantly higher that that of bipedal ($M = 7.80, SD = 1.88$) for standing posture, $t(15) = -3.20, p < .05$. However, there is no significant ($t(15) = -1.32, p = .205$) between unipedal ($M = 9.44, SD = 2.08$) and bipedal ($M = 8.99, SD = 1.26$) for sitting posture on WPM. 

\subsubsection{Error Rate}

RM-ANOVA tests revealed the fact that both posture $(F(1, 15) = 12.73, p < .005, \eta_p^2 = .46)$ and strategies $(F(1, 15) = 8.20, p < .05, \eta_p^2 = .36)$ have significant effect on TER. Post-hoc pairwise comparison showed that the TER of sitting ($M = 3.99, SD = .31$) posture is significantly lower ($p < .005$) than that of a standing posture ($M = 4.88, SD = .33$). In terms of strategies, the TER of bipedal ($M = 3.95, SD = .26$) is significantly lower ($p < .05$) than that of unipedal ($M = 4.93, SD = .41$). 

% overall sitting more accurate than standing
% overall bipedia more accurate than unipedia

Paired-samples t-test indicates no significant effect ($t(15) = -.29, p = .78$) of TER between unipedal ($M = 4.97, SD = 1.71$) and bipedal ($M = 4.79, SD = 1.90$) while standing, but for the sitting posture, the TER of unipedal ($M = 4.89, SD = 1.78$) is significantly higher than that of bipedal ($M = 3.10, SD = 1.12$), $t(15) = 4.48, p < .001$. We then applied Wilcoxon Signed-Rank test on NCER. The results did not show any significant effect between unipedal and bipedal for either standing posture (unipedal: $M = .98, SD = .75$, bipedal: $M = 1.39, SD = 1.50$, $Z = -1.16, p = .245$) or sitting posture (unipedal: $M = 1.06, SD = 1.07$, bipedal: $M = .86, SD = .75$, $Z = -.79, p = .433$). 

% sitting: bipedal more accurate than unipedal
% standing: no error rate difference

\subsubsection{Usability and Workload}

Paired-sample t-test on overall SUS score shows that for standing posture, the score of unipedal ($M = 72.34, SD = 9.55$) is significantly higher than that of bipedal ($M = 64.69, SD = 9.61$), $t(15) = -3.36, p < .005$. However, there is no significant difference between unipedal ($M = 75.47, SD = 12.85$) and bipedal ($M = 72.97, SD = 13.45$) for sitting posture, $t(15) = -1.06, p = .304$. 

% standing: user prefer unipedal than bipedal
% sitting: no preference

We then applied Wilcoxon Signed-Rank test to the NASA-TLX questionnaire results. For the standing posture, participants reported significantly lower mental demand and physical demand when typing with unipedal compared to the bipedal ($p < .005$). They also reported that bipedal input required more effort ($p < .05$), was more frustrating ($p < .05$), and had less perceived performance ($p < .05$) than unipedal input. In contrast, there were no significant differences found between unipedal and bipedal in the sitting posture.

% standing: bipedal high demand, more effort, frustration, less performance
% sitting: no preference

\subsection{Discussion}

% overall unipedal faster than bipedal, but less accurate than bipedal, sitting more accurate than standing

% standing: unipeda faster, no error rate difference, user prefer unipedal than bipedal, bipedal high demand, more effort, frustration, less performance

% sitting: no speed difference, bipedal more accurate than unipedal, user no subjective preference

% Overall speaking, typing with unipedal is generally faster than with bipedal, but less accurate. One possible reason is that for unipedal, navigation and selection are well-integrated into a single foot, which increases the input efficiency. However, such integration will also increase the user's mental and physical demand, leading to a high error rate. In terms of posture, typing in a sitting posture is more accurate than standing posture. One rational explanation is that typing while sitting allows more fine-grained controls, thereby improving accuracy compared to the standing posture.

The results revealed meaningful insights to answer \textbf{RQ4}. 
\revise{ The standing posture requires a higher physical demand which leads to a significantly higher error rate compared with the sitting posture. However,} participants consistently preferred unipedal as it requires less mental and physical demand. Participants commented that when typing using bipedal while standing, they needed to pay much attention to keeping balance, which made them feel tired and tedious. This finding also aligns with \textbf{F5}. \revise{Moreover, typing with unipedal is faster than with bipedal.} Therefore, we consider unipedal is an optimal input strategy for standing posture (\textbf{UPStand}), see Figure~\ref{fig:teaser} \textcircled{a}-\textcircled{d}. 

For sitting posture, there was no significant difference in entry speed between the two strategies, and participants reported no clear preference. However, bipedal typing was significantly more accurate than unipedal one. As bipedal input separates the navigation and selection operation, it also reduces the users' physical demand (See Figure~\ref{fig:Study2_Result.png}\textcircled{e}). Based on these considerations, we selected bipedal as the final input strategy for the sitting posture (\textbf{BPSit}), see Figure~\ref{fig:teaser} \textcircled{e}-\textcircled{h}. 

\section{Study 3: Longitudinal Evaluation}

\revise{Finally, we conducted a 7-day longitudinal user study to examine the progressive learning effect} of \sysName{} in both visual and eyes-free conditions. 
% \textcolor{red}{Specifically, the eye-free condition is defined as entering text without having to pay attention to the keyboard \cite{lu2017blindtype, zhu2019sfree, brewster2003multimodal}.}
\newrevise{
In the literature, eyes-free text entry refers to entering text without having to pay attention to the keyboard \cite{lu2017blindtype, zhu2019sfree, brewster2003multimodal}. 
}
% We were also interested in comparing \sysName's~ performance with current state-of-the-art VR typing techniques. 

\subsection{Participants}

\revise{
We recruited 12 participants (8 male, 4 female) through the university’s internal social media platform, aged between 20-26 years ($M = 22.25, SD = 2.30$). The shoe sizes (EU) of male participants ranged from 40 to 43 ($M = 41.50, SD = 1.20$), and the shoe sizes (EU) of female participants ranged from 37 to 39 ($M = 37.75, SD = .96$). Five participants had prior VR experience, but none reported experiences with VR typing. To ensure a fair comparison, none of the participants had taken part in any of our previous studies. The 12 participants were randomly divided into two groups, with 4 male and 2 female in each group, for two experimental conditions. 
}
% We recruited 6 participants (4 male, 2 female) through the university’s internal social media platform, aged between 20-24 years ($M = 22.50, SD = 1.97$). The shoe sizes (EU) of male participants ranged from 41 to 43 ($M = 42.25, SD = .96$), and those two female participants were 37 and 38, respectively. Three participants had prior VR experience, but none reported experiences with VR typing. To ensure a fair comparison, none of the participants had taken part in any of our previous studies.

\subsection{Apparatus}

We adopted similar hardware and software settings to our previous study (Section~\ref{SEC:layout_apparatus}). \revise{In addition, to support the learnability study under eye-free conditions, we implemented an interface that allows the experiment to be conducted in both \textit{Visual} and \textit{Blind} modes.}

% We adopted similar hardware and software settings to our previous study (Section~\ref{SEC:layout_apparatus}). In addition, to support progressive learning study in eye-free conditions, we implemented an interface to enable switching between full-visual, half-blind, and full-blind modes during runtime. 

\subsection{Study Design and Procedure}

\revise{

We conducted a between-subject study under two typing conditions:
% The two typing conditions we set up for the experiment are as follows: 
1) \textit{Visual.} Participants could see the keyboard layout along with cursor position indicators;
2) \textit{Blind.} Participants could see neither the keyboard layout nor the cursor position indicators \newrevise{\cite{lu2017blindtype, zhu2019sfree, brewster2003multimodal}}. For the study under the \textit{Blind} condition, \newrevise{as suggested by previous works \cite{ye2020qb, wong2018fingert9}, }we provided a cheat sheet mechanism by allowing the users to have a glance at the keyboard layout if they required to. The keyboard layout cue would appear for 10 seconds each time, and we recorded the number of requests each participant made each day. 

% Therefore, whenever participants needed visual cues, the assistant would display the layout information in the VR headset for 10 seconds and logged the number of cues requested by each participant each day.
% Of all participants, half were randomly assigned to the \textit{Visual} condition throughout the experiment, while the other half were \textit{Blind} condition. Except for the visual condition, all participants followed the same experimental procedure.

}

% In order to investigate the fine-grained progressive learning effect for eye-free text entry tasks, our study involved three typing conditions: 
% 1) \textit{Full-Visual.} Participants could see the keyboard layout along with cursor position indicators;
% 2) \textit{Half-Blind.} Participants could see the keyboard layout but without cursor position indicators;
% and 3) \textit{Full-Blind.} Participants could see neither the keyboard layout nor the cursor position indicators. 

\revise{
Each day, participants were required to transcribe one block of 10 phrases using both \textbf{UPStand} and \textbf{BPSit}
% under pre-arranged visual conditions
, with the order counterbalanced across participants. The requirements for standing and sitting postures are the same as in previous experiments (Section~\ref{SEC:stand and seat details}), and the phrases we used every day were picked randomly from MacKenzie's phrase set~\cite{mackenzie2003phrase}.
\revise{We ensured that all selected phrases were case-insensitive and contained no numbers or symbols.}
}
% Each day, participants were required to transcribe two blocks of 10 phrases using both \textbf{UPStand} and \textbf{BPSit}, with the order counterbalanced across participants. One block was performed under the \textit{Full-Visual} condition, and the other under either \textit{Half-Blind} condition or \textit{Full-Blind} condition. The phrases we used every day were picked randomly from MacKenzie's phrase set~\cite{mackenzie2003phrase}.

\revise{
Before the experiment of each day, participants were allowed to practice in either visual or blind condition for as long as they wanted. Throughout the study, participants were instructed to type as quickly and accurately as possible, with a mandatory 5-minute break between blocks to reduce fatigue. Each daily session lasts approximately 45 minutes, and participants were compensated with 40 USD for their time after finishing all sessions. 
}

% To provide consistent training, participants always began each day's study with a block under the \textit{Full-Visual} condition for each posture. The second block varied according to study progress: during the first 4 days, it was conducted under the \textit{Half-Blind} condition, while from the fifth day onward, we removed all visual information and required participants to finish the task in \textit{Full-Blind} condition. Throughout the study, participants were instructed to type as quickly and accurately as possible, with a mandatory 5-minute break between blocks to reduce fatigue. Each daily session lasts approximately 40 minutes, and participants were compensated with 40 USD for their time after finishing all sessions. 

\subsection{Results}
\label{sec:learn_results}

During the study, we measured the entry speed in words per minute (WPM), total error rate (TER), and not corrected error rate (NCER). A Shapiro-Wilk test was performed before the analysis, indicating the WPM and TER  were normally distributed (p > .05), while the NCER 
% of \textbf{BPSit} 
was not normally distributed (p < .05). 
% For a clear illustration, in the following parts of this section, we plot the data of \textit{Half-Blind} condition and \textit{Full-Blind} condition together along the day axis, denoted as \textit{Blind} conditions. 

\begin{figure*} % [h!] is a placement specifier (here, try to place it here)
    \centering % Centers the entire figure
    \begin{subfigure}{0.45\textwidth} % [b] for bottom alignment, 0.45\textwidth sets width
        \centering
        \includegraphics[width=\textwidth]{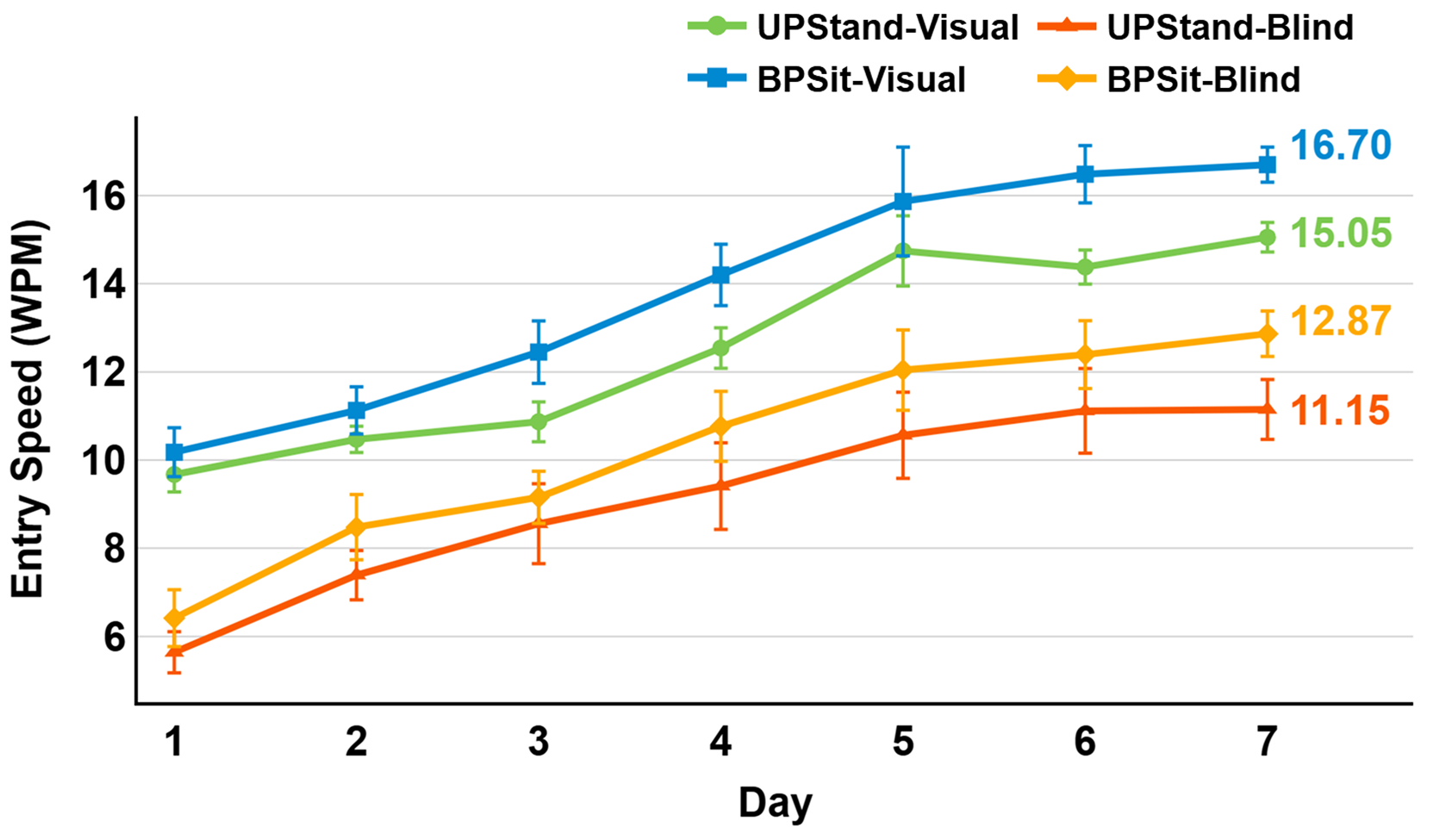} % Include your image
        \caption{Text entry speed (in WPM) across the study period.}
        \label{fig:LC_WPM}
    \end{subfigure}
    % \hfill % Adds horizontal space between subfigures on the same line
    \begin{subfigure}{0.45\textwidth}
        \centering
        \includegraphics[width=\textwidth]{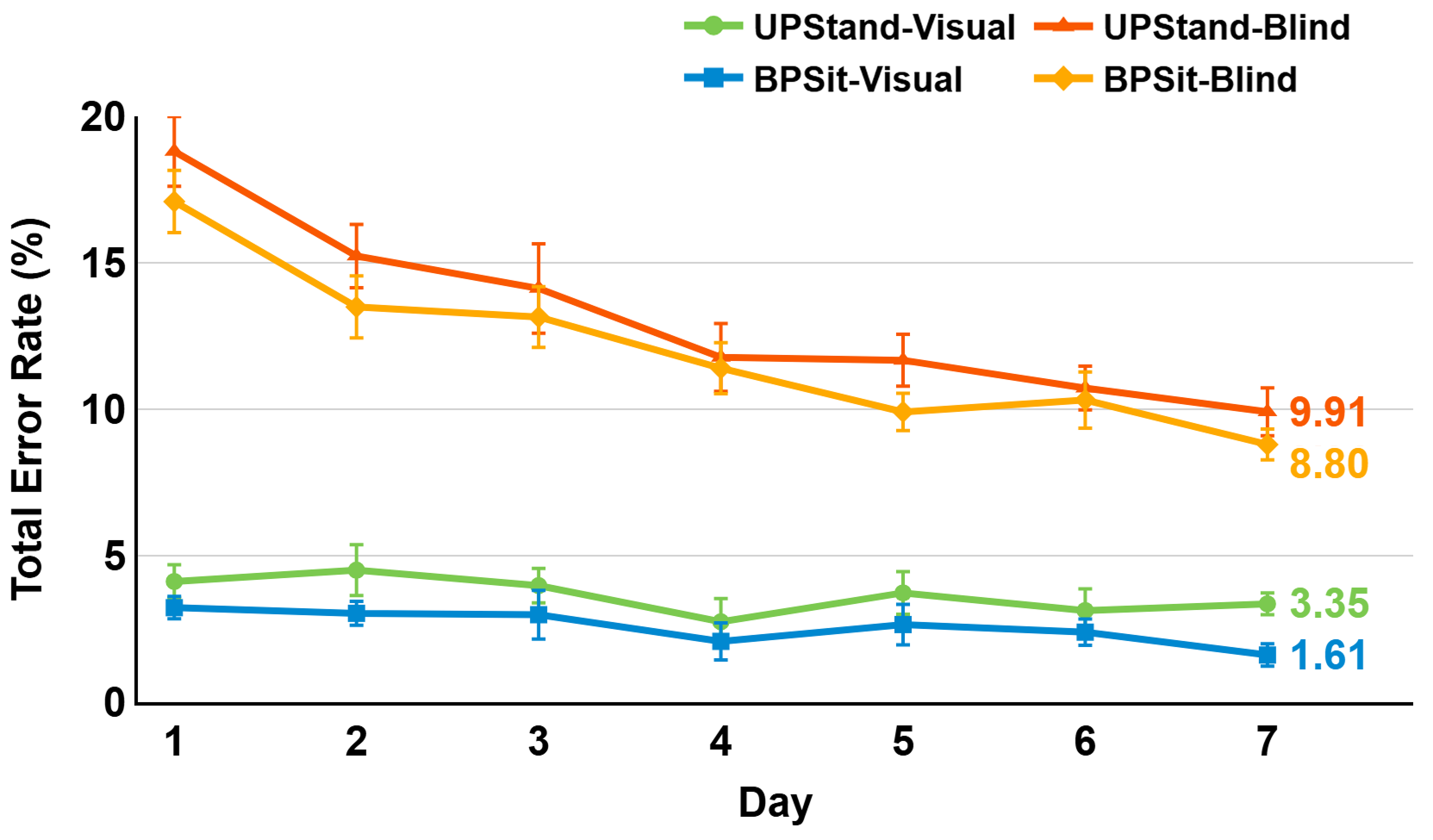}
        \caption{Total error rate across the study period.}
        \label{fig:LC_TER}
    \end{subfigure}
    \begin{subfigure}{0.45\textwidth}
        \centering
        \includegraphics[width=\textwidth]{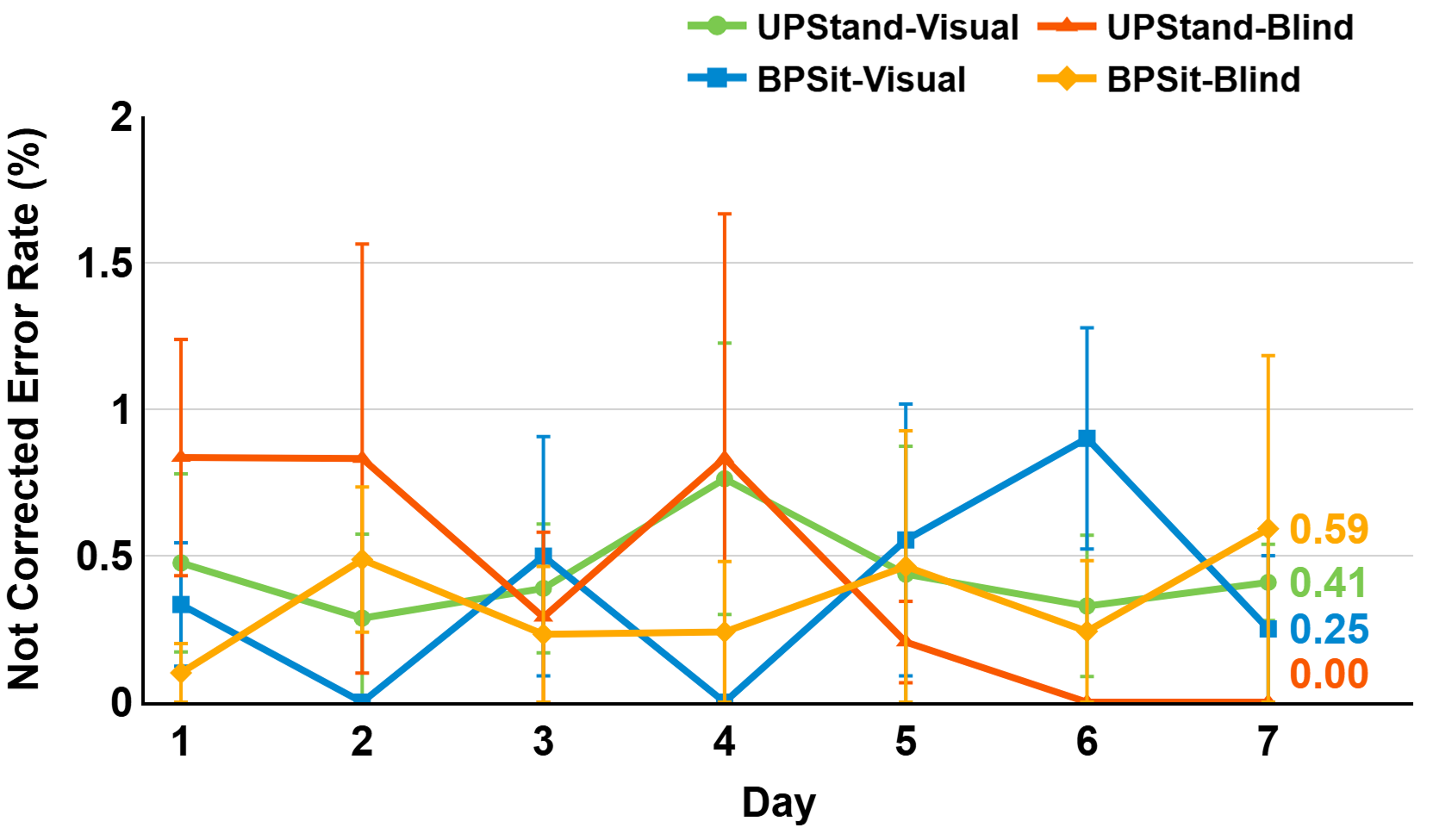}
        \caption{Not corrected error rate across the study period.}
        \label{fig:LC_NER}
    \end{subfigure}
    \Description{Results of the longitudinal user study. (a) Mean of text entry speed (in WPM) across 7 days. (b) Mean of total error rate (TER) across 7 days. (c) Mean of not corrected error rate (NCER) across 7 days. All the following descriptions are in the order of UPStand-Visual, BPSit-Visual, UPStand-Blind, BPSit-Blind. The metrics for the seventh day are as follows: 1. Entry speed is 15.05wpm, 16.70wpm, 11.15wpm and 12.87wpm, respectively. 2. Total error rate is 3.35\%, 1.61\%, 9.91\% and 8.80\%, respectively. 3. Not corrected error rate is 0.41\%, 0.25\%, 0, and 0.59\%, respectively.}
    \caption{Results of the longitudinal user study. (a) Mean of text entry speed (in WPM) across 7 days. (b) Mean of total error rate (TER) across 7 days. (c) Mean of not corrected error rate (NCER) across 7 days. 
    % For all plot above, the data of \textit{Half-Blind} condition and \textit{Full-Blind} condition together along the day axis, denoted as \textit{Blind} condition.
    }
    \label{fig:learning_results}
\end{figure*}

% \begin{figure} 
%     \centering
%     \includegraphics[width=\linewidth]{figs/Learning_Curve.png}
%     \caption{Results of the longitudinal user study. (a) Mean of text entry speed (in WPM) across 7 days. (b) Mean of total error rate (TER) across 7 days. (c) Mean of not corrected error rate (NCER) across 7 days. For all plots above, the data of \textit{Half-Blind} condition and \textit{Full-Blind} condition together along the day axis, denoted as \textit{Blind} condition.}
%     \label{fig:Learning_Curve}
% \end{figure}

\subsubsection{Entry Speed}

% \begin{figure}
%     \centering
%     \includegraphics[width=\linewidth]{figs/learning_wpm.png}
%     \caption{Mean of text entry speed (in WPM) across 7 days. The data of \textit{Half-Blind} condition and \textit{Full-Blind} condition together along the day axis, denoted as \textit{Blind} condition. }
%     \label{fig:learning_wpm}
% \end{figure}

As shown in Figure~\ref{fig:LC_WPM}, all conditions show a clear increasing trend in WPM over time. RM-ANOVA test revealed that day significant effect on WPM for all cases: \textbf{UPStand}-\textit{Visual} $(F(6, 30) = 40.30, p < .005, \eta_p^2 = .89)$, \textbf{BPSit}-\textit{Visual} $(F(6, 30) = 23.01, p < .005, \eta_p^2 = .82)$, \revise{\textbf{UPStand}-\textit{Blind} $(F(6, 30) = 18.04, p < .005, \eta_p^2 = .78)$, and \textbf{BPSit}-\textit{Blind} $(F(6, 30) = 41.32, p < .005, \eta_p^2 = .89)$. }

For the \textit{Visual} condition, the WPM increased from 9.67 to 15.05 (+55.63\%), and from 10.18 to 16.70 for \textbf{BPSit} (+56.07\%), with \textbf{BPSit} slightly outperforming \textbf{UPStand} throughout the period. \revise{We further run post-hoc test on each pair of consecutive days. For \textbf{UPStand}, significant improvements were observed between Day 3 and Day 4 $(p < .001)$, Day 4 and Day 5 $(p < .05)$. For \textbf{BPSit}, post-hoc test revoloves significant improvements between Day 1 and Day 2 $(p < .05)$, Day 2 and Day 3 $(p < .05)$, and Day 3 and Day 4 $(p < .005)$.}

\revise{For the \textit{Blind} conditions, the WPM increased from 5.64 to 11.15 (+97.70\%) for \textbf{UPStand}, and 6.42 to 12.87 (+100.48\%) for \textbf{BPSit}. The post-hoc test shows a significant improvement between the first two days for both \textbf{UPStand} $(p < .001)$ and \textbf{BPSit} $(p < .005)$, and between Day 4 and Day 5 for \textbf{BPSit} $(p < .05)$.}
% For the \textit{Blind} conditions, the WPM increased from 7.19 (Day 1) to 10.05 (Day 4) for \textbf{UPStand}, and 7.63 (Day 1) to 11.33 (Day 4) for \textbf{BPSit}. The value fluctuated on day 5 due to the removal of the visual keyboard layout, but quickly recovered, rising from 9.19 to 11.19 for \textbf{UPStand} and from 10.80 to 12.50 for \textbf{BPSit} by Day 7. Post-hoc tests confirmed significant improvements between Day 5 and Day 7 for both \textbf{UPStand} $(p < .005)$ and \textbf{BPSit} $(p < .05)$. 
These consistently increasing trends of WPM across conditions highlight that participants were able to adapt to \sysName{} quickly. The results also demonstrate the potential of \sysName{} to support efficient text entry, even under challenging eye-free conditions. 

\subsubsection{Error Rate}

As shown in Figure~\ref{fig:LC_TER}, the average TER of \textit{Blind} cases was obviously higher than that of a visual case, but the TER showed a decreasing trend for both cases. For the \textit{Visual} condition, RM-ANOVA did not indicate any significant effect of day. However, TER consistently decreased across the period: from 4.12 to 3.35 for \textbf{UPStand} (-18.68\%), and from 3.23 to 1.61 for \textbf{BPSit} (-50.15\%). In particular, Post-hoc test shows a significant effect $(p < .005)$ for \textbf{BPSit} between Day 1 and Day 7. 

\revise{
In contrast, the \textit{Blind} conditions show stronger learning effects. For \textbf{UPStand}, RM-ANOVA showed a significant effect of time $(F(6, 30) = 27.02, p < .005, \eta_p^2 = .84)$. The TER decreases from 18.80 at Day 1 to 9.91 at Day 7 (-47.29\%). Post-hoc analysis shows significant differences between Day 1 and Day 2 $(p < .001)$ as well as Day 3 and Day 4 $(p < .05)$. For \textbf{BPSit}, RM-ANOVA showed a significant effect of time $(F(6, 30) = 11.73, p < .005, \eta_p^2 = .70)$, with TER decreasing from 17.12 at Day 1 to 8.80 at Day 7 (-48.60\%). Post-hoc test also reveals significant effect for \textbf{BPSit} between Day 1 and Day 2 $(p < .001)$ as well as Day 4 and Day 5 $(p < .05)$. 
}
% In contrast, the \textit{Blind} conditions shows stronger learning effects. For \textbf{UPStand}, RM-ANOVA showed a significant effect of time $(F(6, 30) = 21.53, p < .005, \eta_p^2 = .81)$. The TER decreases from 16.12 (Day 1) to 10.30 (Day 4), and increases to 13.05 (Day 5) after the removal of visual feedback, but then decreases again to 10.08 (Day 7). Post-hoc analysis shows significant differences between Day 5 and Day 7 $(p < .05)$. Similarly, for \textbf{BPSit}, RM-ANOVA showed a significant effect of time $(F(6, 30) = 12.08, p < .005, \eta_p^2 = .70)$, with TER decreasing from 14.41 (Day 1) to 9.95 (Day 4), and increases to 11.95 (Day 5) before decreasing again to 8.63 (Day 7). Post-hoc test also reveals significant effect $(p < .05)$ for \textbf{BPSit} between Day 5 and Day 7. 

\revise{For NCER, Friedman test did not show significant effects across days ($p > .05$). As shown in Figure~\ref{fig:LC_NER}, NCER values fluctuated slightly over days but without clear trends. The overall averages were $0.44$ for \textbf{UPStand}-\textit{Visual}, $0.36$ for \textbf{BPSit}-\textit{Visual}, $0.43$ for \textbf{UPStand}-\textit{Blind}, and $0.34$ for \textbf{BPSit}-\textit{Blind}.}

\revise{

\subsubsection{Number of Layout Cue Requirement}
The average number of keyboard layout cue requirements in the \textit{Blind} condition ranged from 4 to 12 ($M = 7.83, SD = 2.79$) on the first day, with means of 4.67 and 3.83 for \textbf{UPStand} and \textbf{BPSit}, respectively. The average number of requirements on the second day dropped sharply to the range between 0 to 4 ($M = 1.83, SD = 1.72$), with an average of 0.83 times for \textbf{UPStand} and 1 time for \textbf{BPSit}. From the third day onwards, none of the participants required the visual cues.
}

\revise{

\subsection{Discussion}

% Our studies indicate that user's natural ankle spatial awareness is sufficient enough to performance text entry task accurately through training. The longitudinal evaluation further 
The results demonstrate users' potential toward a quicker and more accurate ankle spatial orientation intuition through short-term self-training, both under visual and blind conditions. For the text entry speed,  both \textbf{UPStand} and \textbf{BPSit} under both visual and blind conditions follow similar growth trajectories: the performance improved rapidly in the first 5 days and then gradually stabilized. 
% By the final day, participants achieved an average eye-free performance at approximately 75\% of their average visual typing speed. 
This indicates that users can achieve more precise and faster ankle movement control through short-term practice. It also reveals \sysName's high learning potential under both fully visual and fully blind conditions. Besides, we observed a significant increase in learning effectiveness between day 3 and day 5 under the blind condition. 

The number of keyboard layout cue requirement also highlight a reduction in cognitive effort as users became familiar with \sysName. 
On the first day, participants frequently consulted the keyboard layout, and this number dropped sharply on the second day. From the third day onwards, none of the participants required a keyboard cue at all. This provides an intuition that users are able to quickly memorize the \sysName{} keyboard layout through practice. 

}

\begin{table*}[t]
\begin{tabular}{l||lll|cccc}
\hline
Method & \multicolumn{1}{c}{Year} & \multicolumn{1}{c}{Modality} & \multicolumn{1}{c|}{Affordance} & WPM & TER & NCER & \begin{tabular}[c]{@{}c@{}}Performance after \\ Long-term Train\end{tabular} \\ \hline
RingText~\cite{xu2019ringtext} & 2019 & head motions & hand-free & 12.27 & 3.10\% & 2.25\% & Yes \\ \hline
Blinktype~\cite{lu2020exploration} & 2020 & gaze & hand-free & 13.47 & 10.44\% & 0.90\% & Yes \\ \hline
iText~\cite{lu2021itext} & 2021 & gaze & \begin{tabular}[c]{@{}l@{}}hand-free, \\ invisible keyboard\end{tabular} & 13.77 & \begin{tabular}[c]{@{}c@{}}less than\\ 3\% (WER)\end{tabular} & / & Yes \\ \hline
FeetSymType~\cite{wan2024exploration} & 2024 & foot-based, sitting & hand-free & 11.12 & 3.59\% & 0.46\% & No \\ \hline
SkiMR~\cite{hu2024skimr} & 2024 & gaze & hand-free & 12.05 & / & / & Yes \\ \hline \hline
AnkleType.UPStand & 2025 & foot-based, standing & hand-free & 15.05 & 3.71\% & 0.44\% & Yes \\
AnkleType.BPSit & 2025 & foot-based, sitting & hand-free & \textbf{16.70} & \textbf{2.48\%} & 0.36\% & Yes \\
AnkleType.UPStand & 2025 & foot-based, standing & hand- and eye-free & 11.15 & 9.91\% & 0.43\% & Yes \\
AnkleType.BPSit & 2025 & foot-based, sitting & hand- and eye-free & 12.87 & 8.80\% & \textbf{0.34\%} & Yes \\ \hline
\end{tabular}
\caption{Summary of prior works focusing on hand-free text entry in VR environment. We also show \sysName's~ performance for pair-wise comparison.}
\label{tab:prior_compare}
\end{table*}

\revise{

\section{Application and Discussion}

% \subsection{Performance and Learnability}

% \hl{XXX}

\subsection{Application Scenarios}

\sysName{} was motivated and designed to support situations where users' hands are occupied and less available for text entry. In immersive environments, users' hands may be either engaged with virtual context, such as holding virtual objects or performing direct manipulation gestures with virtual objects, or holding physical objects. In these situations, users' hands become less available to perform conventional text entry. Note that \sysName{} does not target complex dual-task scenarios involving multiple cognitively demanding tasks simultaneously. Instead, we focus on developing an alternative hand- and eye-free text entry approach to enhance the user experiences in virtual environments. The concept of \sysName{} can also be extended to MR/AR scenarios, where users are likely to pay more attention to or interact with a physical artefact. We envision that \sysName{} could be applied to various scenarios. Below, we propose three potential application scenarios across posture and context, as illustrated in Figure~\ref{fig:Scenarios}. 

% \begin{enumerate}
% [label={}, leftmargin=*]
% \item \textcircled{a} A user is browsing and managing content in a VR workspace. He holds a document in one hand, and he inputs virtual sticky notes via \sysName{} without putting the document down. 

% \item \textcircled{b} A user is sitting and watching a VR movie with one hand holding bag of potato chips while the other is engaged in eating. He can input text via \sysName{} to respond to notifications. 

% \item \textcircled{c} A user wearing an MR HMD is walking in a library with books and coffee in both hands. He can quickly respond to short messages using \sysName{} without freeing his hand and leaving his eye off the surroundings. 

% \end{enumerate}

\begin{figure*}
    \centering
    \includegraphics[width=1 \linewidth]{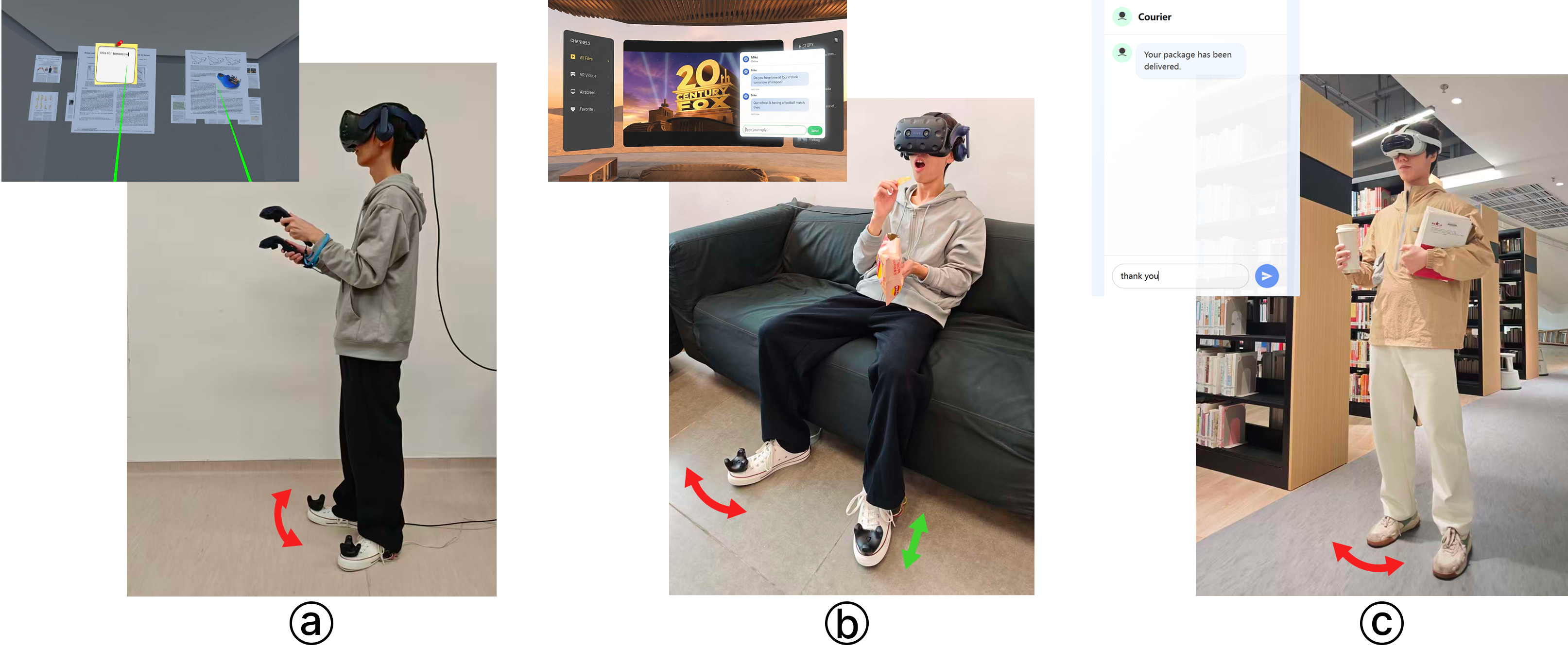}
    \Description{Three potential application scenarios of AnkleType.}
    \caption{\revise{Potential application scenarios of \sysName. \textcircled{a} A user is browsing and managing content in a VR workspace. He grasps and holds two documents using controllers with his left and right hands by pulling the triggers. He can input the virtual sticky notes via \sysName{} without putting the document down. \textcircled{b} A user is sitting and watching a VR movie with one hand holding a bag of potato chips while the other hand is engaged in eating. He can input text via \sysName{} to respond to notifications without clearing his hands. \textcircled{c} A user wearing an MR HMD is walking in a library with books and coffee in both hands. He can quickly respond to short messages using \sysName{} without freeing his hand and leaving his eye off the surroundings.}}
    \label{fig:Scenarios}
\end{figure*}

% - why not gaze 
While gaze-based text entry is intuitive and effective, it occupies visual attention~\cite{velichkovsky1997towards, schwenderling2025teach} and is prone to causing motion sickness~\cite{shi2021virtual,wang2022real,merhi2007motion}. As \sysName{} achieves a competitive performance with other gaze-based techniques (see Table \ref{tab:prior_compare}) without the potential over-use of visual/gaze attention, which is important for many daily tasks. We believe that \sysName{} serves as an alternative input method to gaze in such scenarios.
}

\subsection{Comparison with Prior Works}

We compared the performance of \sysName{} after longitudinal training with prior works on hand-free VR/MR text entry. To the best of our knowledge, no existing research has directly focused on hand- and eye-free text entry in VR/MR. On the other hand, iText~\cite{lu2021itext} presented a gaze-based typing method on an invisible keyboard in VR, which shared a similar design concept with our work. 
%However, there is one inspiring system, iText \cite{lu2021itext}, introduced a gaze-based typing technique on an invisible keyboard, which shares similar design concept with our work. 
As shown in Table~\ref{tab:prior_compare}, \textbf{BPSit} in \textit{Visual} condition outperforms other methods in terms of text entry speed and typing accuracy, reaching the highest entry speed at 16.70 WPM with the lowest TER at 2.48\% \revise{(Section \ref{sec:learn_results})}. More importantly, under the more challenging eye-free condition, both \textbf{UPStand} and \textbf{BPSit} achieve competitive performance compared with prior hand-free approaches. In addition, unlike prior works that primarily focus on a single posture, \sysName{} supports both sitting and standing conditions, with consistent performance across these conditions, enabling a flexible use of \sysName{} across diverse VR scenarios. This comparison shows that \sysName{} not only achieves state-of-the-art performance in the hand-free VR text entry task, but also demonstrates a competitive performance in eye-free scenarios. 

\revise{

\subsection{Foot-based Typing Ergonomics}

% - Source-path-goal
\sysName{} follows a design foundation of heel-pivot rotation with toes as the pointing direction. From an embodied cognition perspective, this design guideline aligns with the Source–Path–Goal image schema \cite{lako1987women, mark1987body}. Humans naturally treat the front direction of the body or limb as the intended direction for moving or targeting. A straight index finger pointing gesture is an intuition of this cognition concept \cite{krause2024perception}. Therefore, heel-pivot rotation and point with toes also align with the body image schema to move from the heel (Source) follow the sole of the foot (Path) to the toes (Goal). This intuition has also been widely adopted in foot-based interaction research \cite{muller2023tictactoes, van2025gesturesock, scott2010sensing}. The user-defined foot gesture yielded from our gesture elicitation study also aligned with this concept: foot flat forward/backward movement naturally maps the toe orientation as the target direction. 

% - Number of joints involved
\sysName{} is highly motivated by recent foot-based text entry research. Compared with prior foot-based text entry systems, our design further minimizes low-limb joint involvement to reduce physical effort. Wan et al. \cite{wan2024exploration} engaged nearly all lower-limb joints (i.e., hip, knee, and ankle) in their typing mechanism and explicitly suggested reducing leg movement in future designs. Building on this recommendation, our design not only mitigates the movement of the hip and knee but also avoids lifting or repositioning of the entire foot. AnkleType mainly relies on ankle movements to perform the main input action, with a small amount of knee movement for menu switching. It does not involve hip movements. Throughout the typing process, AnkleType does not require leaving the entire foot off the ground, which reduces the muscle tension and thus reduces physical demand and fatigue. Involving more joint movements requires more precise control of muscles, which increases fatigue and error-proneness. Moreover, compared with Wan et al. \cite{wan2024exploration}, our design also adapts to standing posture, which further highlights the importance of reducing the number of participating joints when designing foot-based input techniques. 

\subsection{The Speed and Accuracy Trade-Off}

% - The speed and error-rate trade-off
% uni vs. bi
% dominate vs. non-dominate
% alternative use, align with FeetSymType

% Overall speaking, typing with unipedal is generally faster than with bipedal, but less accurate. One possible reason is that for unipedal, navigation and selection are well-integrated into a single foot, which increases the input efficiency. However, such integration will also increase the user's mental and physical demand, leading to a high error rate. In terms of posture, typing in a sitting posture is more accurate than standing posture. One rational explanation is that typing while sitting allows more fine-grained controls, thereby improving accuracy compared to the standing posture.

Our study further highlights a speed-accuracy trade-off in foot-based text entry. Wan et al. \cite{wan2024exploration} explicitly suggested for alternative use of feet to reduce fatigue. Our study extends this perspective and decomposes the typing input into two sub-tasks, namely navigation and control. In the unipedal schema, one foot was used to carry both sub-tasks. This well-integrated interaction task in one foot leads to efficient execution, but it would cause higher fatigue and result in higher error rates. In contrast, the bipedal schema distributes the two sub-tasks across both feet. This subdivision reduces the task load on each foot and thus alleviates user fatigue. But alternative use between the two feet leads to higher task-switching costs, which reduces the typing efficiency but improves the reliability. 

Moreover, our elicitation study reveals that user prefer to assign actions that require higher precision control, such as navigation, to their dominant foot, and leave simple control actions to their non-dominant foot. This asymmetry preference suggests that bipedal design should not simply balance workload across feet, but rather consider natural control capabilities between the dominant and non-dominant foot. 

% \subsection{Technical Feasibility}

% We implemented a proof-of-concept system using customized shoes with tracker and embedded sensors to evaluate AnkleType in VR environment. Beyond VR, the concept of AnkleType has potential and technically feasible to extend to other virtual environment such as AR/MR. For example, light-weighted motion tracker\footnote{https://www.picoxr.com/global/products/pico-motion-tracker} for MR system can achieve similar tracking results. Modern AR/MR HMD such as Apple Vision Pro\footnote{https://support.apple.com/en-us/117810} and Pico 4 Ultra\footnote{https://www.picoxr.com/global/products/pico4-ultra/specs} consist of build-in downward-facing camera array which can also be used to detect foot or ankle movements. These sensing capabilities show potentials to increase the flexibility and mobility of AnkleType to enrich the application scenario. 

}

\section{Limitations and Future Works}

In this section, we highlight the limitations of our research and discuss their potential for future improvements. 

\subsection{Footedness Issue}

The current design of \sysName{} did not account for differences between left- and right-footed users. All participants in our study were right-foot dominant. We draw two potential directions for future work to bridge this gap: 1) explicitly design specific unipedal and bipedal strategies for left-footed users and evaluate, 2) seek adaptive mapping strategies of \sysName{} to support a broader user population. Future work should also investigate how footedness affects performance, user preference, and learning effect.

\subsection{Split-Keyboard Design Space for Bipedal}

The design space of bipedal input strategies remains underexplored. In this work, to maintain design consistency and enable fair pair-wise comparison between unipedal and bipedal strategies, we did not consider involving the left foot for the navigation task. This remains a wider design space in the bipedal ankle-based typing context to provide richer and efficient interaction. One promising future exploration direction is adapting a split-keyboard design, where letters are distributed across both feet. Such a design strategy could better leverage the natural capabilities of bipedal input and expand the keyboard space. With this, we could assign letters more separately, thereby reducing the word ambiguity. At the same time, splitting the keyboard may shorten letter navigation distance, thereby increasing the overall input speed. Thus, we envision great potential in the split-keyboard design, especially for sitting conditions. 

% \textit{Evaluation in Real-Application with Visual Affordence}

% Finally, our evaluation was conducted in a controlled experimental setting. While the eye-free condition in our study demonstrates the potential of \sysName, real-world VR applications often involve complex visual and cognitive attentional demands, such as during gaming, watching movies, or documenting, which may affect users' typing performance. Therefore, future evaluation should be conducted to investigate the use of \sysName{} in more complex VR scenarios.

\subsection{Evaluation in Realistic Dual-Task Scenario}

\sysName{} is motivated by scenarios in which users' hands are physically occupied, such as carrying or holding objects. \newrevise{In our experiments, the participants were not allowed to perform any action with their hands, similar to the constraint of hands cannot move when holding items. We proposed three low–cognitive-load application scenarios (Figure~\ref{fig:Scenarios}) as proof-of-concept demonstrations. However, we did not explicitly evaluate how the primary task in these scenarios affects typing performance, which points to a limitation of our study. 
% In our future work, we will evaluate \sysName's performance under realities applications. 
Moreover, in more complex VR scenarios, such as gaming, users' visual or cognitive attention are mostly occupied by the primarily tasks, which may affect the performance of secondary tasks such as typing. In our longitudinal study, we found that users' typing proficiency and accuracy increased over time, while the number of keyboard layout cue requirements decreased over time as well. This trend suggests that users may develop motor automaticity through long-term practice. Prior researches shows that users' motor automaticity frees up cognitive resources for other processes \cite{beilock2001fragility, poldrack2005neural}, which shows potential to support multiple-task performance \cite{wickens2020processing}. We believe that evaluating \sysName's performance in realistic applications and investigating its potential to support multi-tasking for VR text entry represent promising directions for our future work.}

\subsection{Evaluation in MR context}
% \revise{
% Finally, our study is focused on VR. While the concept of \sysName{} can be potentially extended to MR/AR contexts, as foot movement can be easily captured with the down-facing camera array on current commercial MR/AR HMD. We have not yet conducted experiments in either AR or MR environments, which will be our future work.
% }

\newrevise{We implemented a proof-of-concept system using customized shoes with tracker and embedded sensors to evaluate AnkleType in VR environment. Beyond VR, the concept of AnkleType has potential and technically feasible to extend to other virtual environment such as AR/MR. For example, modern AR/MR HMD such as Apple Vision Pro\footnote{https://support.apple.com/en-us/117810} and Pico 4 Ultra\footnote{https://www.picoxr.com/global/products/pico4-ultra/specs} consist of build-in downward-facing camera array which can also be used to detect foot or ankle movements. These sensing capabilities show potentials to increase the flexibility and mobility of AnkleType to enrich the application scenario. We have not yet conducted experiments in either AR or MR environments, which will be our future work.}

\section{Conclusion}

With \sysName, we primarily contributed a new design space to support foot-based text entry in VR, with a series of user studies and evaluations in this new context. \sysName{} is a novel hand- and eye-free text entry technique for VR that leverages ankle-based gestures to support both standing and sitting postures. Through two preliminary studies, we identified users’ ankle movement ranges and elicited user-preferred gestures, which guide the design of our input strategies. We further optimized the keyboard layout by combining a user study on users’ natural ankle spatial awareness with a computer-simulated language model to reduce word ambiguity. Through a pair-wise comparison user study across 4 user-defined input strategies candidates, we yielded two optimal strategies across postures: \textbf{UPStand} and \textbf{BPSit}, with consideration of both input performance and user preference. Finally, a 7-day longitudinal study demonstrated that participants could achieve promising eye-free typing performance, at \revise{11.15} WPM for \textbf{UPStand} and \revise{12.87} WPM for \textbf{BPSit}, with error rates decreasing over time, achieving a minimum TER of \revise{9.91\%} for \textbf{UPStand} and \revise{8.80\%} for \textbf{BPSit}. Through an offline comparison with previous works, we showed that \sysName{} not only achieves state-of-the-art performance in hand-free VR text entry tasks, but also offers competitive performance in eye-free VR text entry scenarios. 

%%
%% The acknowledgments section is defined using the "acks" environment
%% (and NOT an unnumbered section). This ensures the proper
%% identification of the section in the article metadata, and the
%% consistent spelling of the heading.
\begin{acks}

This research is partially supported by the National
Natural Science Foundation of China, Young Scientists Fund (Project No. 62402301), Natural Science Foundation of Guangdong Province, General Research Fund (Project No. 2025A1515010236), and the STU Scientific Research Initiation Grant (SRIG, Project No. NTF23024). This research is also partially supported by the Centre for Applied Computing and Interactive Media (ACIM) of School of Creative Media, City University of Hong Kong. 

\end{acks}

%%
%% The next two lines define the bibliography style to be used, and
%% the bibliography file.
\bibliographystyle{ACM-Reference-Format}
\bibliography{sample-base}

%%
%% If your work has an appendix, this is the place to put it.
% \appendix

% \section{Research Methods}

% \subsection{Part One}

% Lorem ipsum dolor sit amet, consectetur adipiscing elit. Morbi
% malesuada, quam in pulvinar varius, metus nunc fermentum urna, id
% sollicitudin purus odio sit amet enim. Aliquam ullamcorper eu ipsum
% vel mollis. Curabitur quis dictum nisl. Phasellus vel semper risus, et
% lacinia dolor. Integer ultricies commodo sem nec semper.

% \subsection{Part Two}

% Etiam commodo feugiat nisl pulvinar pellentesque. Etiam auctor sodales
% ligula, non varius nibh pulvinar semper. Suspendisse nec lectus non
% ipsum convallis congue hendrerit vitae sapien. Donec at laoreet
% eros. Vivamus non purus placerat, scelerisque diam eu, cursus
% ante. Etiam aliquam tortor auctor efficitur mattis.

% \section{Online Resources}

% Nam id fermentum dui. Suspendisse sagittis tortor a nulla mollis, in
% pulvinar ex pretium. Sed interdum orci quis metus euismod, et sagittis
% enim maximus. Vestibulum gravida massa ut felis suscipit
% congue. Quisque mattis elit a risus ultrices commodo venenatis eget
% dui. Etiam sagittis eleifend elementum.

% Nam interdum magna at lectus dignissim, ac dignissim lorem
% rhoncus. Maecenas eu arcu ac neque placerat aliquam. Nunc pulvinar
% massa et mattis lacinia.

\end{document}